\documentclass[a4paper,11pt]{article}
\usepackage{jheppub}
\usepackage{eurosym}
\usepackage{amssymb}
\usepackage{amsfonts,cite}
\usepackage{amsmath}
\usepackage{graphicx}
\usepackage{multirow}
\usepackage{xcolor}
\usepackage{epsfig}
\usepackage{fancyhdr}
\usepackage{float}

\newbox\mybox

\newcommand\fverb{\setbox\mybox=\hbox\bgroup\verb}
\newcommand\fverbdo{\egroup\medskip\noindent\fbox{\unhbox\mybox}\ }
\newcommand\fverbit{\egroup\item[\fbox{\unhbox\mybox}]}

\abstract{We investigate Hamiltonian curl-force systems with indefinite kinetic energy. We first reconsider Berry's polynomial Hamiltonian curl-force model, whose numerically observed closed trajectories motivated an integrability conjecture. A Painlev\'e analysis yields a non-principal resonance spectrum, so that the corresponding Laurent series cannot accommodate the required number of arbitrary constants of the general solution. The model therefore fails the standard Painlev\'e test. We then introduce a four-parameter curl-force family and identify the parameter locus on which this system is integrable. We construct a second Hamiltonian, compatible Poisson tensors, separated complex characteristic variables, and a Lax representation. More generally, the separated form yields polynomial integrable curl-force Hamiltonians of arbitrary degree. We also show that the same construction admits a higher time-derivative potentialisation whose free limit is the degenerate Pais-Uhlenbeck oscillator. Finally, we analyse zero-curl invariant reductions and elliptic periodic solutions, and exhibit an isolated periodic orbit outside the integrable regime. This illustrates that closed trajectories alone do not imply Liouville or Painlev\'e integrability.}  

\title{Integrable curl-force Hamiltonians: bi-Hamiltonian structure, separability, and periodic orbits}

 \author[a]{Alexander Felski,}
 \author[b]{Andreas Fring}

 \emailAdd{alexander.felski.d8@tohoku.ac.jp}
 \emailAdd{a.fring@city.ac.uk}	
 

\affiliation[a]{Advanced Institute for Materials Research, Tohoku University, Sendai, Japan}
 \affiliation[b]{Department of Mathematics, City St George's, University of London,  Northampton Square, \\ London EC1V 0HB, UK}

\begin{document}
 	\maketitle
 	
 	\pagestyle{fancy}
 	\fancyhead{} 
 	\fancyhead[LE,RO]{\small\itshape Integrable curl-force Hamiltonians} 
 	
 	\renewcommand{\headrulewidth}{0.4pt}
 	
 \section{Introduction}
 
 Curl forces, introduced and studied systematically by Berry and Shukla, are
 position-dependent Newtonian forces whose curl does not vanish and which
 therefore cannot be written as ordinary Euclidean gradient forces
 \cite{berry2012class}. They are nonconservative, but need not be
 dissipative in the sense that the corresponding flow in position-velocity space is
 volume-preserving. Physical realisations arise, for example, in the motion
 of dipoles near optical vortices \cite{berry2013phys}, the description of odd elasticity \cite{scheibner2020odd,chen2021realization,gao2022non,zhang2024anisotropic}, 
 odd viscosity \cite{avron1995viscosity,avron1998odd,fruchart2023odd,caprini2025odd,marini2026emer}, 
  Lorentz-force-like and lift-force effects \cite{lapa2014swimming,kogan2016lift,lou2022odd,lier2023lift},
 current-induced atomic motion in molecular junctions and atomic nanowires
 \cite{dundas2009current,bode2011scattering,lu2012current},
  and in nano-optomechanics and optical tweezers \cite{roichman2008influence,wu2009direct,gloppe2014bidimensional}.
 
 Although generic
 curl forces are not Hamiltonian in the usual sense, Berry and Shukla showed
 that a large class can be generated by Hamiltonians with anisotropic
 quadratic kinetic energy \cite{berry2015hamiltonian}. When this kinetic metric is indefinite, 
 as in the models studied here, Hamiltonian curl forces provide a natural bridge to
 ghostly Hamiltonians and, through them, to higher time-derivative
 theories, as noted by Guha \cite{guha2020curl}. This connection is particularly natural for Pais-Uhlenbeck type models,
 which provide the standard finite-dimensional testing ground for higher
 time-derivative dynamics \cite{pais1950field,smilga2005benign}.
 
 The Hamiltonian formulation of curl forces also raises the natural question of whether
 these theories can be consistently quantised. Within the constant-metric natural class, such Hamiltonians require a non-Euclidean kinetic metric, which in the present model is
 indefinite. Hence, their quantisation of the present indefinite-metric models is expected to
 encounter the familiar difficulties of ghostly systems, i.e. unbounded energy
 operators, continuous spectra, generalised eigenfunctions, and the possible
 absence of a conventional normalisable ground state. This is precisely the
 type of phenomenon found in recent attempts to quantise Hamiltonian curl
 forces. The quantisation of the simplest Hamiltonian curl
 force \cite{berry2024quant} leads to an unbounded Hamiltonian with continuous-spectrum stationary
 states and unusual singularity structure. Similarly, complex Hamiltonian
 curl-force systems, while classically integrable and separable in complex
 variables, lead after quantisation to unfamiliar wavefunctions \cite{berry2020class}. These
 features make Hamiltonian curl forces a natural testing ground for the same
 issues that arise in ghostly Hamiltonians and, ultimately, in second-order
 representations of higher time-derivative theories. Recent work has also shown how Lie symmetries, bi-Hamiltonian structures
 and non-unitary similarity transformations can be used to construct
 ghost-free or normalisable representations of such systems
 \cite{FFT,felski2026three,fring2026ghost}.
 
 The integrability of curl-force systems is another subtle question. Berry and
 Shukla studied symmetries, constants of motion and nondissipative chaotic
 behaviour, introducing numerical diagnostics to distinguish hidden
 invariants from apparently invariant-free dynamics
 \cite{berry2016curl}. In a related direction, Berry constructed a class
 of complex Hamiltonian curl forces which are completely integrable and
 separable in complex variables \cite{berry2020class}. More recently, a hierarchy
 of curl-force examples was presented, ranging from Hamiltonian integrable
 curl forces to non-Hamiltonian systems with no conserved quantities
 \cite{berry2025six}. These works make clear that the existence of regular or
 closed trajectories is not, by itself, a reliable integrability criterion.
 
 In this work we revisit one of the Hamiltonian curl-force examples
 motivated by this programme. We show that the conjectured integrability of Berry's polynomial
 Hamiltonian curl-force model does not hold in the standard
 Painlev\'e sense: all admissible dominant balances are
 non-principal and fail to accommodate the required four arbitrary constants
 of the general solution. We then construct an integrable polynomial modification of this model. On the integrable locus the same condition is
 obtained from several independent viewpoints: the Painlevé test, the
 existence of a second Hamiltonian, a compatible bi-Hamiltonian structure,
 separation in complex characteristic variables, and a separated Lax
 representation.
 
Our paper is organised as follows: In section 2 we recall the relevant
 properties of curl forces and derive a criterion for constant-metric
 Hamiltonian realisations, thereby making explicit the relation to ghostly
 Hamiltonians. In section 3 we revisit Berry's polynomial curl-force Hamiltonian
 and show that it fails the standard Painlev\'e test. In section 4 we introduce
 an integrable modofication of this model and demonstrate that the same parameter condition
 underlies the Painlev\'e compatibility, the second Hamiltonian, the
 bi-Hamiltonian formulation, the separation of variables and the Lax pair.
 In section 5 we analyse the curl geometry and its implications for periodic
 trajectories, including invariant zero-curl reductions, explicit elliptic-function
 solutions and an isolated periodic orbit outside the integrable regime. In
 section 6 we discuss the higher time-derivative potentialisation and the
 degenerate Pais-Uhlenbeck limit. In appendix A we provide a general polynomial
 construction, which produces integrable curl-force Hamiltonians of arbitrary
 degree.
 
 \section{Hamiltonian curl forces and ghostly models}
 
 \subsection{General properties of curl forces}
 
 Curl forces are a special class of Newtonian forces $\ddot{\bf r} = {\bf F} ({\bf r})$ whose curl does not vanish identically $\nabla \times {\bf F} \neq 0$. In two dimensions we
 embed the planar force field as $ \mathbf F(x,y)=(F_x(x,y),F_y(x,y),0)$,
 its three-dimensional curl is
 \begin{equation}
 	\nabla\times\mathbf F
 	=
 	\left(\partial_xF_y-\partial_yF_x\right)\mathbf e_z .
 \end{equation}
 In what follows we use the standard two-dimensional convention and denote
 the scalar $z$-component by  $\nabla\times\mathbf F:=\partial_xF_y-\partial_yF_x$.
Some general properties relevant for the class considered here are:
 \begin{itemize}
\item[i)] In general curl forces are nonconservative, meaning that they cannot be expressed as a gradient force in the ordinary Euclidean sense, i.e. they cannot be written as  $ \mathbf F=-\nabla \Phi$, with $\Phi$ being a scalar field. 
 \item[ii)]  Thus the work $W_\Gamma$  around a closed loop need not vanish. For a closed
 curve $\Gamma$ bounding a region $\Sigma$, Stokes' theorem gives
 \begin{equation}
 	W_\Gamma
 	=
 	\oint_\Gamma \mathbf F\cdot d\mathbf r =
 	\iint_\Sigma
 	(\nabla\times\mathbf F)\,dx\,dy ,     \label{Stokeswork}
 \end{equation}
 which is generically nonzero, although it may vanish for special
 curves because of cancellation of the signed curl flux.
 
 \item[iii)]  The phase-space flow is incompressible so that the dynamics is nondissipative. For a two-dimensional system this is seen as follows. Writing the second-order equations as a first-order system by introducing $ v_x:=\dot x$, $v_y:=\dot y$ we have
 \begin{equation}
 	\frac{d}{dt}
 	\begin{pmatrix}
 		x\\
 		y\\
 		v_x\\
 		v_y
 	\end{pmatrix}
 	=
 	X(x,y,v_x,v_y)
 	:=
 	\begin{pmatrix}
 		v_x\\
 		v_y\\
 		F_x(x,y)\\
 		F_y(x,y)
 	\end{pmatrix}.
 \end{equation}
 The phase-space divergence of this vector field is
 \begin{equation}
 	\nabla_{\rm ph}\cdot X	=	\frac{\partial v_x}{\partial x}	+
 	\frac{\partial v_y}{\partial y}	+	\frac{\partial F_x}{\partial v_x}	+	\frac{\partial F_y}{\partial v_y}.     \label{phasediv}
 \end{equation}
 Here $x,y,v_x,v_y$ are regarded as independent phase-space coordinates.
 Moreover, for the class of autonomous position-dependent curl forces considered
 here, $F_x$ and $F_y$ do not depend on the velocities. Therefore each term in (\ref{phasediv}) vanishes, so that
 \begin{equation}
 	\nabla_{\rm ph}\cdot X=0.
 \end{equation}
 This condition means that the phase-space flow is incompressible. 
 
\item[iv)] Any autonomous Newtonian system of the form
 \begin{equation}
 	\ddot{\mathbf q}=\mathbf F(\mathbf q),
 \end{equation}
 with a force depending only on the configuration variables, possesses
 the reversing transformation
 \begin{equation}
 	\mathcal R:(\mathbf q,\dot{\mathbf q}) \longmapsto (\mathbf q,-\dot{\mathbf q}). \label{reversetrans}
 \end{equation}
 This statement is independent of whether the force is conservative or
 has nonvanishing curl. In particular, if $\mathbf q(t)$ is a
 solution, then $\mathbf q(-t)$ is also a solution. The two solutions
 need not coincide. However, an individual trajectory is invariant under this
 reversal when its initial state is chosen in the fixed-point set
 of $\mathcal R$, namely when $\dot{\mathbf q}=0$.
 
  \end{itemize}
 	
 \subsection{Constant-metric Hamiltonian realisations}
 	
 	A non-vanishing Euclidean curl excludes a conventional natural
 	Hamiltonian formulation with Euclidean kinetic energy, but it does
 	not exclude more general Hamiltonian formulations. In particular,
 	the system may admit a natural Hamiltonian realisation with a
 	non-Euclidean and possibly indefinite kinetic metric. We now derive
 	a simple criterion for the existence of such a constant-metric
 	Hamiltonian realisation.
 	
 	Considering  a general dynamical system of the form
 	\begin{equation}
 		\ddot q^i=F^i(q),
 	\end{equation}
 	and assuming that it is generated by a canonical natural Hamiltonian
 	\begin{equation}
 		H(q,p)
 		=
 		\frac{1}{2}p_iG^{ij}p_j+V(q),
 	\end{equation}
 	where $G$ is a constant real symmetric nondegenerate, $\det G\neq0$, kinetic
 	metric. Then Hamilton's equations imply
 	\begin{equation}
 		\ddot{\boldsymbol q}
 		=
 		-G\nabla V.
 	\end{equation}
 	Writing $M=G^{-1}$, the potential must therefore satisfy
 	\begin{equation}
 		\nabla V=-M\boldsymbol F.
 	\end{equation}
 	The existence of such a potential requires the corresponding mixed
 	partial derivatives to agree,
 	\begin{equation}
 		\frac{\partial}{\partial q^k}
 		\left(M_{ij}F^j\right)
 		=
 		\frac{\partial}{\partial q^i}
 		\left(M_{kj}F^j\right).
 	\end{equation}
 	Since $M$ is constant, this condition can be written compactly as
 	\begin{equation}
 		MDF=(DF)^TM,
 		\label{critcurlcond}
 	\end{equation}
 	where $DF$ denotes the Jacobian matrix of the force field. This is
 	the condition that $M$ symmetrise the force Jacobian.
 	
 	For the Euclidean kinetic metric one has $M=I$, in which case
 	\eqref{critcurlcond} reduces to $DF=(DF)^T$.
 	Thus, a non-vanishing Euclidean curl implies that the identity matrix
 	cannot symmetrise the force Jacobian and therefore excludes a
 	conventional natural Hamiltonian with Euclidean kinetic energy.
 	A Hamiltonian formulation may nevertheless exist if there is another
 	constant real symmetric nondegenerate matrix $M$ satisfying
 	\eqref{critcurlcond}. In that case $M$ may be interpreted as the inverse
 	kinetic metric, and $-M\boldsymbol F$ is locally the gradient of a
 	scalar potential even though $\boldsymbol F$ itself is not a
 	Euclidean gradient.
 	
 	In two dimensions, writing
 	\begin{equation}
 		M=
 		\begin{pmatrix}
 			a&c\\
 			c&b
 		\end{pmatrix},
 		\qquad
 		\boldsymbol F=
 		\begin{pmatrix}
 			F_x\\
 			F_y
 		\end{pmatrix},
 	\end{equation}
 	the compatibility condition reduces to
 	\begin{equation}
 		a\,\partial_yF_x
 		+c\left(\partial_yF_y-\partial_xF_x\right)
 		-b\,\partial_xF_y=0.
 		\label{curlcond2D}
 	\end{equation}
 	Since \eqref{curlcond2D} is homogeneous in $a$, $b$ and $c$, a
 	definite Hamiltonian realisation exists if it admits a solution with $ ab-c^2>0$,
 	as the overall sign of $M$ may then be chosen so that $M$ is positive
 	definite. If every nondegenerate solution instead satisfies $ab-c^2<0$,
 	then every constant-metric natural Hamiltonian realisation has
 	Lorentzian signature and is therefore ghostly.
 	
 	The criterion \eqref{critcurlcond} is restricted to natural Hamiltonians
 	with a constant kinetic metric. More general Hamiltonian
 	realisations involving coordinate-dependent metrics or noncanonical
 	Poisson structures are not excluded when no such matrix $M$ exists.
 	
 	\section{Berry's curl-force Hamiltonian and Painlev\'e  obstruction }
 	
 	We consider the Hamiltonian \cite{berry2015hamiltonian,berry2025six}
 	\begin{equation}
 		H(x,y,p_x,p_y)	=	\frac{1}{2}\left(p_x^2-p_y^2\right)	+	\frac{1}{4}\left(x^4-y^4\right)	+	xy^2-yx^2 . \label{Ham1}
 	\end{equation}
 	Then Hamilton's equations are
 	\begin{eqnarray}
 		\dot{x} &=&\frac{\partial H}{\partial p_x} =p_x , 	\qquad \,\, \,\,	\dot{p}_x=-\frac{\partial H}{\partial x}=-x^3-y^2+2xy, \\
 		\dot{y}&=&\frac{\partial H}{\partial p_y}=-p_y,	\qquad
 		\dot{p}_y=-\frac{\partial H}{\partial y} = y^3-2xy+x^2.
 	\end{eqnarray}
 	Combining these equations gives the two coupled second-order differential equations
 	\begin{equation}
 		\ddot{x}=-x^3-y^2+2xy =-V_x, \qquad  	\ddot{y}=-y^3+2xy-x^2 = V_y, \label{twosec}
 	\end{equation}
 	where $V(x,y)$ is the potential to be read off from (\ref{Ham1}).
 	
 	This system admits a non-vanishing force field
 	\begin{equation}
 		\mathbf F(x,y)=\bigl(F_x,F_y\bigr)=\bigl(-V_x,V_y\bigr),
 	\end{equation}
 	with non-vanishing curl
 	\begin{equation}
 		\nabla\times\mathbf F	=	4(y-x) .
 	\end{equation}
 
Based on the numerical observations that identified closed trajectories it was conjectured that it is also integrable. In Section 5 we will show that exploiting the reversibility property one can construct isolated periodic trajectories even outside the integrable parameter regime. We now also show explicitly that the integrability conjecture cannot hold in the standard Painlev\'e sense \cite{Pain1,Pain2,ramani1989painleve}. 

We consider now the system (\ref{twosec}). Following the standard procedure we introduce
\begin{equation}
	\tau=t-t_0,
\end{equation}
with $t_0$ being a movable singularity and assume the leading-order behaviour to be of the form
\begin{equation}
	x\sim a_0\tau^p,
	\qquad
	y\sim b_0\tau^q.   \label{movable}
\end{equation}
We observe that the most singular terms are the cubic terms, so that
\begin{equation}
	\ddot{x}\sim -x^3,
	\qquad
	\ddot{y}\sim -y^3.
\end{equation}
Thus balancing the powers gives
\begin{equation}
	p=q=-1.
\end{equation}
From substitution into the dominant equations we get
\begin{equation}
	2a_0=-a_0^3,
	\qquad
	2b_0=-b_0^3,
\end{equation}
so that
\begin{equation}
	a_0=\pm i\sqrt{2},
	\qquad
	b_0=\pm i\sqrt{2}.   \label{a0boberry}
\end{equation}
Perturbing the leading behaviour now as
\begin{equation}
	x=a_0\tau^{-1}+\phi\tau^{r-1},	\qquad  	y=b_0\tau^{-1}+\psi\tau^{r-1}.
\end{equation}
Substituting into the equations of motion (\ref{twosec}), linearising in $\phi, \psi$ the order $\tau^{r-3}$ yields 
\begin{equation}
	(r-1)(r-2)\phi	=	-3 a_0^2\phi	\qquad 
	(r-1)(r-2)\psi	=	-3 b_0^2\psi.  \label{domb}
\end{equation}
Using the solutions from (\ref{a0boberry}) both equations reduce to
\begin{equation}
	(r-1)(r-2)=6, \qquad \Rightarrow \qquad r=-1,\;4.
\end{equation}
Since the system consists of two second-order equations, its general
solution must contain four arbitrary constants. A principal dominant
balance should therefore possess the universal resonance $r=-1$,
associated with the movable singularity $t_0$, together with three
further admissible resonances at which the remaining arbitrary
constants enter. Instead, for each of the dominant branches in
\eqref{a0boberry}, we obtain
\begin{equation}
	r=-1,-1,4,4.
\end{equation}
The repeated negative resonance shows that these balances are
non-principal. Even if the double resonance at $r=4$ introduces two
independent arbitrary coefficients, the corresponding right Laurent
series does not contain the four arbitrary constants required for
the general solution. Since no principal dominant balance exists,
the system fails the standard Painlev\'e test and is not integrable in the strong Painlevé sense. 
Failure of the standard Painlev\'e test does not by itself exclude
Liouville integrability, integrability on special invariant
manifolds, or integrability after a nontrivial transformation, see e.g. \cite{ramani2000integrable}. 

\section{A polynomial curl-force family with an integrable locus}

We now propose an extended version of (\ref{Ham1}) that does pass the Painlev\'e test. We consider the Hamiltonian
\begin{equation}
	H_I(x,y,p_x,p_y)	=	\frac{1}{2}\left(p_x^2-p_y^2\right)	+ \frac{\alpha}{2} x^2 + \frac{\beta}{2} y^2  + \mu \left(	xy^2-yx^2  \right) +
		 \nu    \left(x^4-y^4 	 + 4 xy^3- 4 y x^3  \right) . \label{Hamint1}
\end{equation}
with equations of motion
\begin{eqnarray}
	\ddot{x}	&=&  -\alpha  x -\mu  \left(y^2-2 x y\right)  -\nu  \left(4 x^3-12 x^2 y+4 y^3\right) =: F_x ,   \label{nHqunom1} \\
		\ddot{y} &=& \beta  y	 - \mu  \left(x^2-2 x y\right)  -  \nu  \left(4 x^3- 12 x y^2+4 y^3\right) =:F_y   .  \label{nHqunom2}
\end{eqnarray}
Notice that $H(x,y,p_x,p_y)	$ in (\ref{Ham1}) cannot be obtained from $H_I(x,y,p_x,p_y)$ in some simple limit. As the Hamiltonian in the previous section, also this Hamiltonian admits a curl force interpretation. With
$\mathbf F(x,y)=\bigl(F_x,F_y\bigr)=\bigl(-V_x,V_y\bigr)$
 we obtain now
\begin{equation}
	\nabla\times\mathbf F	=	4(y-x)\bigl[\mu+6\nu(x+y)\bigr].  \label{curlFint}
\end{equation}
Thus, the force is generically a curl force with properties as stated in section 2. Next we show that this system is integrable. 

\subsection{Painlev\'e test}
We assume again the leading-order behaviour to be of the form (\ref{movable}).
The dominant terms near the singularity are the cubic nonlinear terms, so we take
\begin{equation}
	p=q=-1.
\end{equation}
Substituting into (\ref{nHqunom1}) and (\ref{nHqunom2}) the dominant parts of the equations gives
\begin{equation}
	2a_0+4\nu\left(a_0^3-3a_0^2b_0+b_0^3\right)=0,   \qquad
2b_0+ 4\nu\left(a_0^3-3a_0b_0^2+b_0^3\right) =0.
\end{equation}
Introducing $s:=b_0/a_0 $ we replace $b_0$ in both equations, then multiplying the first equation by $s/(2 a_0) $ and the second by $1/(2 a_0) $ and subsequently taking  the difference of the two equations yields 
\begin{equation}
	2 \nu a_0^2 (s-1)(s+1)(s^2 -s +1) =0 .
\end{equation}
Thus, we have the three sets of solutions 
\begin{equation}
	s =1,  \,\,a_0^2 = \frac{1}{2 \nu}  ,  \qquad \,\,
	s = -1,    \,\,  a_0^2 =- \frac{1}{6 \nu}  ,   \qquad\,\,
	s =   \frac{1\pm i\sqrt{3}}{2} ,    a_0^2 = \frac{1}{6 \nu s}.  \label{loesung}
\end{equation}
Next we perturb the leading-order solutions
\begin{equation}
	x 	= 	a_0\tau^{-1} 	+	\phi\tau^{r-1},   	\qquad	y	=	b_0\tau^{-1}	+ \psi\tau^{r-1},
\end{equation}
substituting into the equations of motion (\ref{nHqunom1}) and (\ref{nHqunom2}), linearising in $\phi, \psi$ the order $\tau^{r-3}$ yields the two equations
\begin{equation}
	\begin{pmatrix}
		\rho+12 a_0^2 \nu -24 a_0 b_0 \nu   & 12 \nu  \left(b_0^2-a_0^2\right) \\
		12 \nu  \left(a_0^2-b_0^2\right) & \rho +12 b_0^2 \nu  - 24 a_0 b_0 \nu 
	\end{pmatrix}
	\begin{pmatrix}
		\phi\\
		\psi
	\end{pmatrix}
	=
	\begin{pmatrix}
		0\\
		0
	\end{pmatrix},
\end{equation}
with $\rho =(r-1)(r-2)$. Thus,  for a nontrivial perturbation, the determinant of the coefficient matrix must vanish. Together with the first two solutions for $a_0$, $b_0$ in (\ref{loesung}) we only get $\rho=6$ as solutions, and hence not enough resonances for the Painlev\'e tests to pass. However,  the two complex principal branches
$s=(1\pm i\sqrt{3})/2$, together with the two choices of $a_0$,
yield $\rho=0$ and $\rho=6$, corresponding to
\begin{equation}
	r=-1,1,2,4.     \label{resonances}
\end{equation}
We find the expected resonance at $r=-1$ and three more, which equal the required number in this case for Painlev\'e tests to pass.
 The resonances $r=1,2,4$ must be checked for compatibility.

For this purpose we consider  the full Laurent expansions
\begin{equation}
	x=\sum_{j=0}^{\infty}a_j\tau^{j-1},
	\qquad
	y=\sum_{j=0}^{\infty}b_j\tau^{j-1}.   \label{Laurent}
\end{equation}
Then at each order $j$, the recursion relations take on the form
\begin{equation}
	A_j
	\begin{pmatrix}
		a_j\\
		b_j
	\end{pmatrix}
	=
	C_j,
\end{equation}
for some $2 \times 2$ matrix $A$ and the vector $C_j$ depending only on lower-order coefficients.  In detail we have:

Starting with the Laurent expansion (\ref{Laurent}) terminated at $j=1$, taking the last solution in (\ref{loesung}) and $s$ to be the solution with the plus sign, the equations of motion (\ref{nHqunom1}) and (\ref{nHqunom2}) can be written as
\begin{equation}
	\frac{2}{s}	
	 \begin{pmatrix}
	1-2 s & s-2 \\
	s-2 & 1+s 
	\end{pmatrix}    \begin{pmatrix}
	a_1\\
	b_1
	\end{pmatrix}     = \frac{\mu}{6 \nu s}   \begin{pmatrix}
	1+s\\
	1-2 s
	\end{pmatrix}   .
\end{equation}
An arbitrary constant $c_1$ enters at this point by noting that the matrix on the left hand side possesses a null vector $(s,1)^\intercal$. We can therefore add it  to $(a_1,b_1)^\intercal$ multiplied by a constant that we denote by $c_1$. Solving thereafter for the unknown coefficients gives
\begin{equation}
	   a_1 = -\frac{\mu}{12 \nu s}  + c_1 s, \qquad b_1 =c_1 .
\end{equation} 
Terminating next at $j=2$ converts the equations of motion into 
\begin{equation}
\frac{2}{s}	
\begin{pmatrix}
	1-2 s & s-2 \\
	s-2 & 1+s 
\end{pmatrix} 
   \begin{pmatrix}
	a_2\\
	b_2
\end{pmatrix}     =- \frac{1}{\sqrt{ 6 s \nu  } }   \begin{pmatrix}
	 \alpha  + \frac{\mu^2}{ 12 \nu}	\\
		 s \left(  \beta  -  \frac{\mu^2}{ 12 \nu}  \right)
	\end{pmatrix}   .
\end{equation}
The coefficient matrix again possesses the null vector
$(s,1)^{\mathrm T}$. Compatibility of the right-hand side requires
$\alpha=-\beta$, after which the general solution is
\begin{equation}
	a_2=  i s \sqrt{\frac{2}{  \nu s}}   \left( \frac{\beta}{12}  -\frac{ \mu^2}{144 \nu }   \right)         + c_2s, \qquad b_2 =c_2 .
\end{equation} 
At the next level $j=3$ we do not expect to find a resonance according to (\ref{resonances}). Indeed, we compute
\begin{equation}
	2 \begin{pmatrix}
		-s & 1-\frac{2}{s} \\
		1-\frac{2}{s} & s^* 
	\end{pmatrix} 
	\begin{pmatrix}
		a_3\\
		b_3
	\end{pmatrix}  =  
	\begin{pmatrix}    
	  \frac{\frac{\mu ^3}{36 \nu }+ s^* \beta  \mu }{12 \nu }+\left(\frac{(2 s^* +1) \mu ^2}{12 \nu
	  }-s \beta \right) c_1+3 s^* \mu  c_1^2+12 s^*\nu  c_1^3    \\
	  \frac{(3 s-1) \mu ^3}{432 \nu ^2}+\left(\beta +\frac{(3 s-1) \mu ^2}{12 \nu }\right) c_1+3 s \mu 
	  c_1^2+12 s \nu  c_1^3
	\end{pmatrix}.
\end{equation}
Since the coefficient matrix on the left-hand side is nonsingular at
$j=3$, the coefficients $a_3,b_3$ are uniquely determined by the lower
orders. Terminating next at $j=4$ gives the equations of motion in the form  
\begin{equation}
	2
	\begin{pmatrix}
		2- s & 1 - \frac{2}{s} \\
		1 - \frac{2}{s} & -1-s 
	\end{pmatrix} 
	\begin{pmatrix}
		a_4\\
		b_4
	\end{pmatrix}     =  C_4,
\end{equation}
where $C_4$ is a complicated non-vanishing vector whose explicit form we do not report. Crucially,
we notice that the matrix on the left is singular so that we may once again add its null vector, $(1/s,1)^\intercal$ in this case, multiplied by an arbitrary constant $c_4$. We verified that this equation may be solved for $a_4$ and $b_4$ with no additional constraints.

Therefore the arbitrary constants in the principal Laurent expansion are
\begin{equation}
    t_0, \qquad c_1, \qquad  c_2, \qquad c_4, \quad \text{for} \qquad \alpha =- \beta .
\end{equation}
Thus, the compatibility check does indeed produce the four arbitrary constants, but the Painlev\'e test is only passed with the additional constraint $\alpha =- \beta$ on the model parameters.

\subsection{Bi-Hamiltonian structure}

We now show that the integrable case admits a bi-Hamiltonian formulation, i.e. we can find two Hamiltonians $H_1$, $H_2$ and two Poisson tensors $J_1$, $J_2$, that produce the same flow 
\begin{equation}
	\dot z	=	J_1\nabla H_1	=	J_2\nabla H_2 ,     \label{biflow}
\end{equation}
where $
z=(x,y,p_x,p_y)^T $. For $\beta=-\alpha$, the Hamiltonian $H_I$ from (\ref{Hamint1}) becomes
\begin{equation}
	H_1=
	\frac12(p_x^2-p_y^2)
	+\frac{\alpha}{2}(x^2-y^2)
	+\mu(xy^2-yx^2)
	+\nu(x^4-y^4+4xy^3-4yx^3).    \label{Hamint1b}  
\end{equation}
As the second Hamiltonian we identify
\begin{eqnarray}
	H_2(x,y,p_x,p_y)	&=& \frac{1}{2} \left(p_x^2+p_y^2 \right)  + 2 p_x p_y +\frac{\alpha}{2}   \left(x^2-4
	x y+y^2\right) \qquad \,	\label{Hamint2}  \\
	&&-\frac{ \mu}{3}  \left(2 x^3-3 x^2 y-3 x y^2+2 y^3\right)  -\nu  \left(x^4+4 x^3 y-12 x^2 y^2+4 x y^3+y^4\right) . \notag
\end{eqnarray}
Defining the two constant skew-symmetric matrices
\begin{equation}
	J_1 =  \begin{pmatrix}
		0 & 0 & 1 & 0 \\
		0 & 0 & 0 & 1 \\
		-1 & 0 & 0 & 0 \\
		0 & -1 & 0 & 0 
	\end{pmatrix}, \qquad   J_2 =  \frac{1}{3}  \begin{pmatrix}
		0 & 0 & -1 & 2 \\
		0 & 0 & -2 & 1 \\
		1 & 2 & 0 & 0 \\
		-2 & -1 & 0 & 0   
	\end{pmatrix} ,
\end{equation}
we can verify (\ref{biflow}). Here the matrix $J_1$ is the canonical Poisson tensor, and since $J_2$ is also
constant and skew-symmetric, it defines a second Poisson bracket. The Jacobi identities are satisfied trivially. Moreover,
any linear combination
\begin{equation}
J_\lambda=J_2-\lambda J_1
\end{equation}
is again a constant skew-symmetric Poisson tensor. Thus $J_1$ and $J_2$
are compatible Poisson structures. By standard arguments it also follows that $H_2$ is conserved along the $H_1$-flow and vice versa.

\subsection{Separability}
We now show that both Hamiltonians $H_1$ and $H_2$ are separable. In general we can write the two Hamiltonians as
\begin{eqnarray}
		H_1(x,y,p_x,p_y)	&=&	\frac{1}{2}\left(p_x^2-p_y^2\right)	+ V(x,y)  \label{H1V}\\
   	H_2(x,y,p_x,p_y)	&=& \frac{1}{2} \left(p_x^2+p_y^2 \right)  + 2 p_x p_y + W(x,y)  \label{H2W}
\end{eqnarray}
where the two potentials $V(x,y)$ and $W(x,y)$ are defined in an obvious way by comparing with (\ref{Hamint1}) and (\ref{Hamint2}), respectively. Computing now the Poisson brackets
\begin{equation}
	\left\{ H_1,H_2  \right\}_1 =\sum_{z=x,y} \frac{\partial H_1}{ \partial z} \frac{\partial H_2}{ \partial p_z} - 
	\frac{\partial H_1}{ \partial p_z} \frac{\partial H_2}{ \partial z} =
	p_x\left( V_x + 2 V_y  - W_x    \right)  + p_y \left(  V_y+ 2 V_x + W_y   \right) .  \label{VWequn}
\end{equation}
For this to vanish both coefficients on the right-hand side need to be zero. This is indeed the case for our system and hence $H_1$ and $H_2$ are in involution.

We can, however, analyse this further. Differentiating the first bracket with respect to $y$ and the second with respect to $x$ and subsequently adding the two results we obtain
\begin{equation}
	  0= V_{xx} + V_{xy} + V_{yy} = \alpha + \beta ,   \label{Vequn}
\end{equation}
which only holds in the integrable case. Next we notice that the operator acting on $V$ factorises
\begin{equation}
\partial_x^2 + \partial_x \partial_y + \partial_y^2
= (\partial_x - \rho \partial_y)(\partial_x - \rho^2 \partial_y),  \label{factoriseop}
\end{equation}
provided $\rho^3 =1$ and $\rho^2 + \rho + 1=0$, i.e. $\rho = e^{2 \pi i/3}$. Thus, the characteristic variables for this operator are 
\begin{equation}
       u := y + \rho x, \quad v:= y + \bar{\rho} x ,  \quad  \Leftrightarrow   \quad
       x= \frac{u-v}{\rho - \bar{\rho}}, \quad  y= \frac{\rho u -v}{\rho-1}      \label{uvsep}
\end{equation}
and therefore the potential is separable
\begin{equation}
	 V(x,y) = F(u)  + G(v) .  \label{Vxyuv}
\end{equation}
In appendix \ref{polynomialpot} we provide a general polynomial solution for $V(x,y)$ and $W(x,y)$.

The corresponding canonical momenta are obtained by demanding the canonical one-form to be preserved
\begin{equation}
     p_x dx + p_y dy = p_u du + p_v dv. 
\end{equation}
Solving this equation we find
\begin{equation}
	p_x = \rho p_u + \bar{\rho} p_v, \quad p_y = p_u + p_v , \quad  \Leftrightarrow   \quad
	p_u = \frac{p_x - \bar{\rho} p_y}{\rho- \bar{\rho}}, \quad  	p_v = \frac{p_x - \rho p_y}{\bar{\rho} - \rho }. \label{pupvsep}
\end{equation}
Trading the canonical variables $x,y,p_x,p_y$ in the Hamiltonians for the new set of canonical variables $u,v,p_u,p_v$ they both separate into two disjoint anharmonic oscillators
\begin{eqnarray}
H_1(u,v,p_u,p_v)&=& h_u(u,p_u) + h_v(v,p_v), \quad \label{huhv1} \\
H_2(u,v,p_u,p_v) &=&  i \sqrt{3} \left[   h_v(v,p_v) - h_u(u,p_u)  \right].  \label{huhv2}
\end{eqnarray}
With $\alpha=-\beta=\omega^2$, where $\omega^2\in\mathbb{R}$ is not
assumed to be positive, the separated equations become
\begin{eqnarray}
   h_u(u,p_u) & =&  -\frac{1}{2} (\rho +2) p_u^2+\frac{1}{6} (\rho -1) \omega ^2 u^2-\frac{1}{9} (2 \rho
   +1) \mu  u^3-\frac{1}{3} (\rho +2) \nu  u^4 ,\\
   h_v(v,p_v) & =&   -\frac{1}{2} (\bar{\rho} +2) p_v^2+\frac{1}{6} (\bar{\rho} -1)\omega ^2 v^2-\frac{1}{9} (2 \bar{\rho}
   +1) \mu 
   v^3-\frac{1}{3} (\bar{\rho} +2) \nu  v^4.
\end{eqnarray}
The equations of motion in the separated variables follow directly from the canonical Hamilton equations
\begin{eqnarray}
	\dot u &=&\frac{\partial H_1}{\partial p_u}= -(\rho+2)p_u,    \label{udpu} \\
	\dot p_u&=&-\frac{\partial H_1}{\partial u}= 	-\frac13(\rho-1)\omega^2u
	+\frac13(2\rho+1)\mu u^2
	+\frac43(\rho+2)\nu u^3,\\
	\dot v&=&\frac{\partial H_1}{\partial p_v}= -(\bar{\rho}+2)p_v ,   \label{vdpv}   \\
	\dot p_v&=&-\frac{\partial H_1}{\partial v}= 	-\frac13(\bar{\rho}-1)\omega^2v
	+\frac13(2\bar{\rho}+1)\mu v^2
	+\frac43(\bar{\rho}+2)\nu v^3.
\end{eqnarray}
Eliminating the momenta gives two decoupled second-order equations
\begin{equation}
		\begin{aligned}
			\ddot u+\omega^2u+\rho\mu u^2-4 \bar{\rho}\nu u^3&=0,\\
			\ddot v+\omega^2v+\bar{\rho}\mu v^2-4\rho\nu v^3&=0.    
		\end{aligned} \label{sepeqom}
\end{equation}
For real $x,y,\mu,\nu$ the variables $u$ and $v$ are complex conjugates,
$ v=\bar u$, so that the two separated equations are complex conjugates of one another.

\subsection{Lax pair representation}

Having established separability, we now construct a Lax representation
adapted to the separated variables \cite{Lax,OP6,OP2}. Setting the Hamiltonians to constants
\begin{equation}
	H_1=E,\qquad H_2=I,   \label{Hconst}
\end{equation}
we obtain with (\ref{huhv1}) and (\ref{huhv2}) the two separated constants
\begin{equation}
	\varepsilon_u:=h_u	= 	\frac12\left(E+\frac{i}{\sqrt3}I\right),
	\qquad
	\varepsilon_v:=h_v 	= 	\frac12\left(E-\frac{i}{\sqrt3}I\right).
\end{equation}
Thus the two independent conserved quantities may be represented either by
$(H_1,H_2)$ or by $(h_u,h_v)$. Using the relation (\ref{udpu}), (\ref{vdpv}) we eliminate $p_u$, $p_v$ from the equations $h_u=\varepsilon_u$, $h_v=\varepsilon_v$ and obtain the
first-order separated equations
\begin{equation}
	\dot u^2=\mathcal P_u(u),
	\qquad
	\dot v^2=\mathcal P_v(v),   \label{uveqoms}
\end{equation}
where
\begin{align}
	\mathcal P_u(\lambda)
	={}&
	2\bar\rho\nu\lambda^4
	-\frac23\rho\mu\lambda^3
	-\omega^2\lambda^2
	-2(\rho+2)\varepsilon_u,
	\\
	\mathcal P_v(\lambda)
	={}&
	2\rho\nu\lambda^4
	-\frac23\bar\rho\mu\lambda^3
	-\omega^2\lambda^2
	-2(\bar\rho+2)\varepsilon_v .
\end{align}
Here the variable $\lambda$ is introduced as an auxiliary coordinate on
the algebraic curve associated with the separated motion. The
separated $u$-equation in (\ref{uveqoms}) is expressed equivalently as
$ \eta^2=\frac14\mathcal P_u(u)$,
if we define $\eta:=\frac12\dot u$. This is the energy curve of the one-dimensional separated problem. To
describe the whole curve, rather than just the moving point on it, we
replace the dynamical coordinate $u(t)$ by a free coordinate
$\lambda$. Thus we consider the algebraic curve
\begin{equation}
	\eta^2=\frac14\mathcal P_u(\lambda).   \label{algcurv}
\end{equation}
The pair $(\lambda,\eta)$ are coordinates on this curve: $\lambda$ is
the horizontal coordinate and $\eta$ is the corresponding value of the
square root. The actual solution $u(t)$ determines a moving point on the
curve by
\begin{equation}
\lambda=u(t),
\qquad
\eta=\frac12\dot u(t).
\end{equation}
This curve becomes the spectral curve once it is realised as the
characteristic equation of the Lax matrix. Namely, we construct
$L_u(\lambda)$ so that
\begin{equation}
\det \left[ \eta \mathbb{I}  -  L_u(\lambda)  \right]=0   \label{detetaL}
\end{equation}
is precisely the algebraic curve (\ref{algcurv}). Thus $\lambda$ is the spectral parameter labelling the family of matrices $L_u(\lambda)$, while $\eta$ is the eigenvalue of $L_u(\lambda)$. 

We now construct $L_u(\lambda)$ so that its characteristic equation
coincides with this curve. Making the generic Ansatz for a traceless $2\times2$ matrix,
\begin{equation}
L_u(\lambda)
=
\begin{pmatrix}
	A_u(\lambda) & B_u(\lambda)\\
	C_u(\lambda) & -A_u(\lambda)
\end{pmatrix}.
\end{equation}
Then equation (\ref{detetaL}) yields
\begin{equation}
\eta^2-\left(A_u^2+B_uC_u\right) =0 .
\end{equation}
We want this to match equation (\ref{algcurv}) so that we have
\begin{equation}
A_u^2+B_uC_u=\frac14\mathcal P_u(\lambda).
\end{equation}
A natural choice is
\begin{equation}
B_u(\lambda)=\lambda-u,
\qquad
A_u(\lambda)=\frac12\dot u.
\end{equation}
The zero of $B_u$ records the current position of the separated
coordinate. At $\lambda=u$, the eigenvalue on the spectral curve is
$\eta=\dot u/2$. Using $\dot u^2=\mathcal P_u(u)$, the remaining entry
is then fixed
\begin{equation}
C_u(\lambda)
=
\frac{\mathcal P_u(\lambda)-\mathcal P_u(u)}
{4(\lambda-u)}.
\end{equation}
Hence
\begin{equation}
L_u(\lambda)
=
\begin{pmatrix}
	\dfrac12\dot u
	&
	\lambda-u
	\\[2mm]
	\dfrac{\mathcal P_u(\lambda)-\mathcal P_u(u)}
	{4(\lambda-u)}
	&
	-\dfrac12\dot u
\end{pmatrix}.      \label{Lumat}
\end{equation}
Since $\mathcal P_u(\lambda)-\mathcal P_u(u)$ is divisible by
$\lambda-u$, this matrix is regular at $\lambda=u$.

The corresponding $M_u$-matrix is obtained by requiring the Lax equation \cite{Lax}
\begin{equation}
	\dot L_u=[M_u,L_u].   \label{Laxequ}
\end{equation}
We find
\begin{equation}
		M_u(\lambda)
		=
		\begin{pmatrix}
			0&1\\[2mm]
			\dfrac{
				\mathcal P_u(\lambda)-\mathcal P_u(u)
				-(\lambda-u)\mathcal P_u'(u)
			}
			{4(\lambda-u)^2}
			&0
		\end{pmatrix}.   \label{Mumat}
\end{equation}
The numerator has a double zero at $\lambda=u$, so $M_u(\lambda)$ is
also regular at $\lambda=u$. The Lax equation (\ref{Laxequ})
is then equivalent to
\begin{equation}
	\dot u^2=\mathcal P_u(u),
	\qquad
	\ddot u=\frac12\mathcal P_u'(u),
\end{equation}
and hence to the separated equation of motion in (\ref{sepeqom}).

The $v$-sector is obtained by the simultaneous replacements
\begin{equation}
u\to v,\qquad
\mathcal P_u\to \mathcal P_v,\qquad
\varepsilon_u\to\varepsilon_v,\qquad
\rho\leftrightarrow\bar\rho .
\end{equation}
We find that $L_v(\lambda)$ and $M_v(\lambda)$ are the same matrices as in (\ref{Lumat}) and (\ref{Mumat}) with $u \rightarrow v$. Then
\begin{equation}
	\dot L_v=[M_v,L_v], \quad  \Leftrightarrow \quad 
	\dot v^2=\mathcal P_v(v),	\,\,\, 	\ddot v=\frac12\mathcal P_v'(v), \quad \Leftrightarrow \quad 
	\ddot v+\omega^2v+\bar\rho\mu v^2-4\rho\nu v^3=0.
\end{equation}
The full Lax pair is the direct sum of the two separated blocks
\begin{equation}
		L(\lambda)
		=
		\begin{pmatrix}
			L_u(\lambda)&0\\
			0&L_v(\lambda)
		\end{pmatrix},
		\qquad
		M(\lambda)
		=
		\begin{pmatrix}
			M_u(\lambda)&0\\
			0&M_v(\lambda)
		\end{pmatrix},
\end{equation}
satisfying the Lax equation
\begin{equation}
		\dot L(\lambda)=[M(\lambda),L(\lambda)].
\end{equation}

Since the eigenvalues of $L(\lambda)$ are conserved in time, equation (\ref{algcurv}) implies that the coefficients of the polynomials $P_u(\lambda)$ and $P_v(\lambda)$ are also time-independent. Indeed, the coefficients of $\lambda^4,\lambda^3,\lambda^2$ are fixed parameters
of the model, while the constant terms are proportional to $\varepsilon_u$, $\varepsilon_v$, which are the conserved energies. 
Moreover, the Lax equation implies that $ \operatorname{tr}L^{k}$ is conserved for any positive integer $k$. In particular we find for these trace invariants
\begin{equation}
	\operatorname{tr}L_z^{2n}
	=
	2\left(\frac{\mathcal P_z(\lambda)}{4}\right)^n,
	\qquad
	\operatorname{tr}L_z^{2n+1}=0,    \quad z=u,v  .
\end{equation}
Thus, for the full block-diagonal Lax matrix we have
\begin{equation}
	\operatorname{tr}L^{2n}
	=
	2\left(\frac{\mathcal P_u(\lambda)}{4}\right)^n
	+
	2\left(\frac{\mathcal P_v(\lambda)}{4}\right)^n,
	\qquad
	\operatorname{tr}L^{2n+1}=0.
\end{equation}
In particular, for $n=1$, using $\rho+\bar\rho=-1$, we obtain
\begin{equation}
	\operatorname{tr}L^2
	=
	-\nu\lambda^4
	+\frac13\mu\lambda^3
	-\omega^2\lambda^2
	-\frac32H_1
	+\frac12H_2 .
\end{equation}
Thus $H_1$ and $H_2$ appear as coefficients of the spectral invariant.
In this normalisation the Hamiltonian is not equal to the whole expression
$\operatorname{tr}L^2$, as is often the case, rather, the Hamiltonians are encoded in its
spectral coefficients.

The separated Lax pair can be pulled back to the original canonical
variables by using the inverse of the canonical transformation in (\ref{uvsep}) and (\ref{pupvsep}).
In this way one obtains a Lax pair representation entirely in
$(x,y,p_x,p_y)$. It remains, however, the pull-back of the separated Lax
pair in which its block structure reflects the separation in the
complex characteristic variables $u$ and $v$. 

\section{Curl geometry and periodic trajectories}

Although the force is nonconservative in the Euclidean configuration
plane, every periodic Newtonian trajectory satisfies a useful
constraint. Let $\Gamma$ denote the configuration-space projection of
a periodic orbit of period $T$. Since $\mathbf F(x,y)=(\ddot x,\ddot y)$,
the circulation of the force along $\Gamma$ is
\begin{equation}
	\oint_\Gamma \mathbf F\cdot d\mathbf r  =
	\int_0^T     \mathbf F\cdot  \dot{ \mathbf  r}     \,dt
	=
	\int_0^T
	(\ddot x\dot x+\ddot y\dot y)\,dt
	=
	\frac12[\dot x^2+\dot y^2]_0^T=0.
\end{equation}
Here we have used the periodicity of the full phase-space orbit,
$ \dot x(T)=\dot x(0)$, $\dot y(T)=\dot y(0)$. If $\Gamma$ is a positively oriented simple closed curve bounding a
region $\Sigma$, Green's theorem gives
\begin{equation}
	\iint_\Sigma
	\left(\partial_xF_y-\partial_yF_x\right)
	\,dx\,dy
	=
	\iint_\Sigma
	(\nabla\times\mathbf F)\,dx\,dy
	=0.
	\label{periodic-curl-flux}
\end{equation}
Thus, the signed curl flux through the region enclosed by a simple
periodic trajectory must vanish. For a self-intersecting projection, the corresponding statement
involves the winding-number-weighted signed curl flux.

Consequently, a simple periodic orbit cannot enclose a region of definite
curl sign. Its interior must have vanishing signed curl flux. 
For the integrable system we have the curl of the force given by (\ref{curlFint}), so that the zero-curl
lines
\begin{equation}
y=x,\qquad \mu+6\nu(x+y)=0   \label{zerocurl}
\end{equation}
provide a natural geometric skeleton for the configuration-space motion.

Next we address the question under which conditions a zero-curl line can support a trajectory
which remains on it. We consider the equations of motion in the form (\ref{nHqunom1}) and (\ref{nHqunom2}). 
For a straight line
\begin{equation}
\ell=\{(x,y): n_1x+n_2y=c\},
\end{equation}
with normal vector $n=(n_1,n_2)$, a trajectory remains on $\ell$ only if
three conditions are satisfied. First, the initial point must lie on the
line
\begin{equation}
n_1x(0)+n_2y(0)=c.
\end{equation}
Second, the initial velocity must be tangent to the line
\begin{equation}
n_1\dot x(0)+n_2\dot y(0)=0.
\end{equation}
Third, the force must be tangent to the line along the line itself
\begin{equation}
n_1F_x(x,y)+n_2F_y(x,y)=0,
\qquad (x,y)\in \ell .
\end{equation}
The first two are conditions on the initial data, while the last one is a
condition on the vector field. Without the last condition the trajectory
would leave the line due to a nonzero normal acceleration.

For the first line in (\ref{zerocurl}) with $n=(1,-1)$ and $c=0$  the conditions  become
\begin{equation}
  x(0)=  y(0), \qquad \dot x(0)= \dot y(0), \qquad  \alpha= - \beta.  \label{zerocurl1}
\end{equation}

For the second line in (\ref{zerocurl}) with $n=(1,1)$ and $c=- \mu/ 6 \nu $  the conditions become
\begin{equation}
x(0)+y(0)=-\frac{\mu}{6\nu}, \qquad \dot x(0)= - \dot y(0), \qquad  \alpha= - \beta= -\frac{\mu^2}{18\nu}. \label{zerocurl2}
\end{equation}
We will implement these initial conditions and constraints in our numerical analysis that follows.

\subsection{Nonperiodic growing motion}

\begin{figure}[H]
	\begin{minipage}[b]{\textwidth}      
		\centering
		\includegraphics[width=0.49\textwidth]{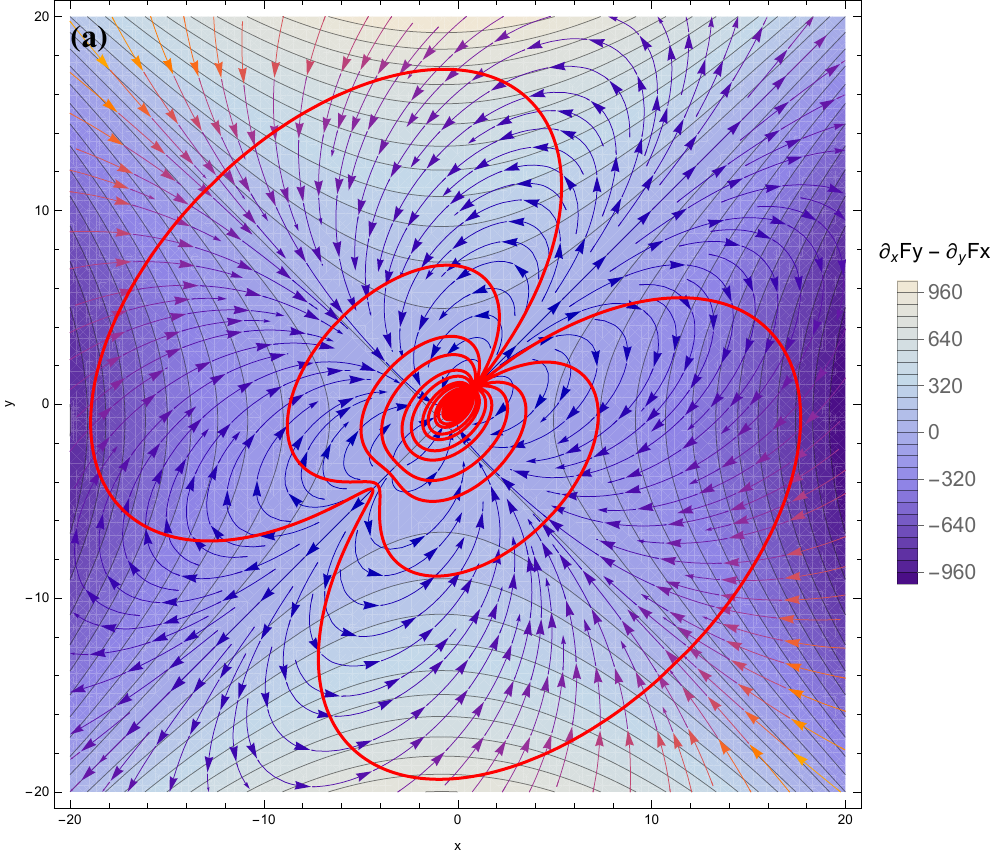}
		\includegraphics[width=0.49\textwidth]{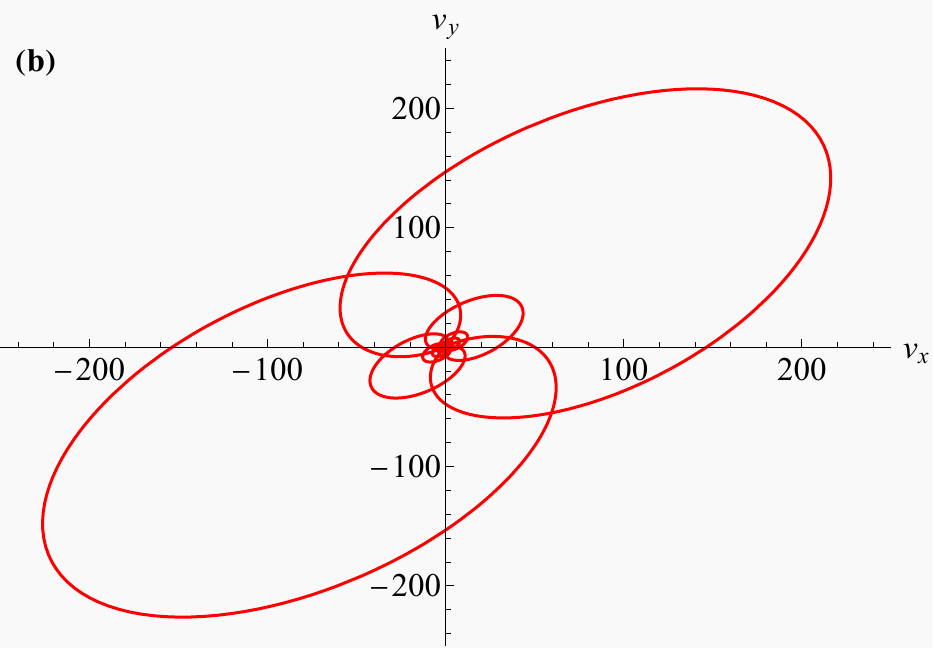}
		\includegraphics[width=0.49\textwidth]{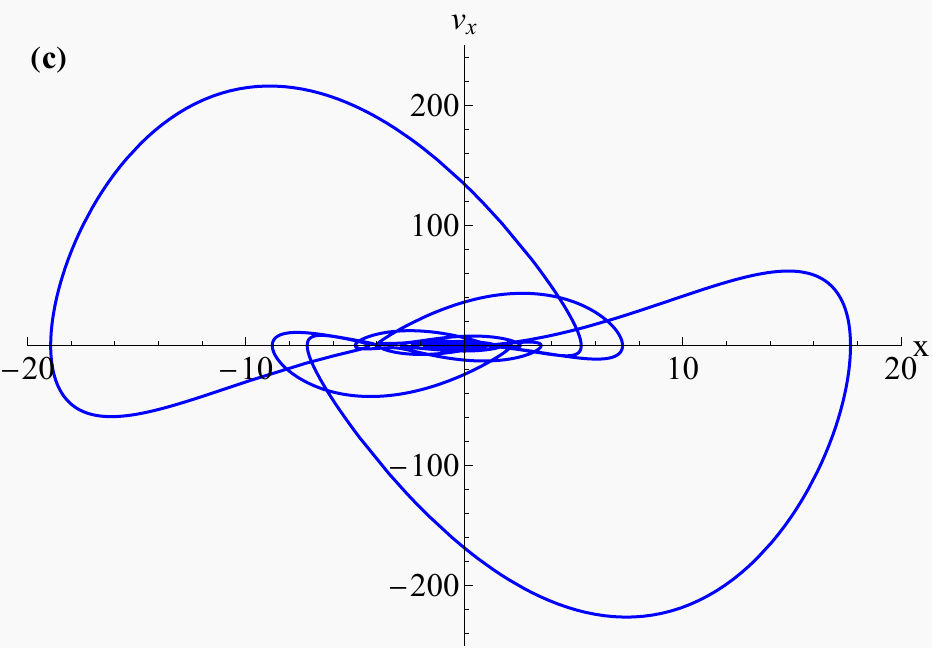}
		\includegraphics[width=0.49\textwidth]{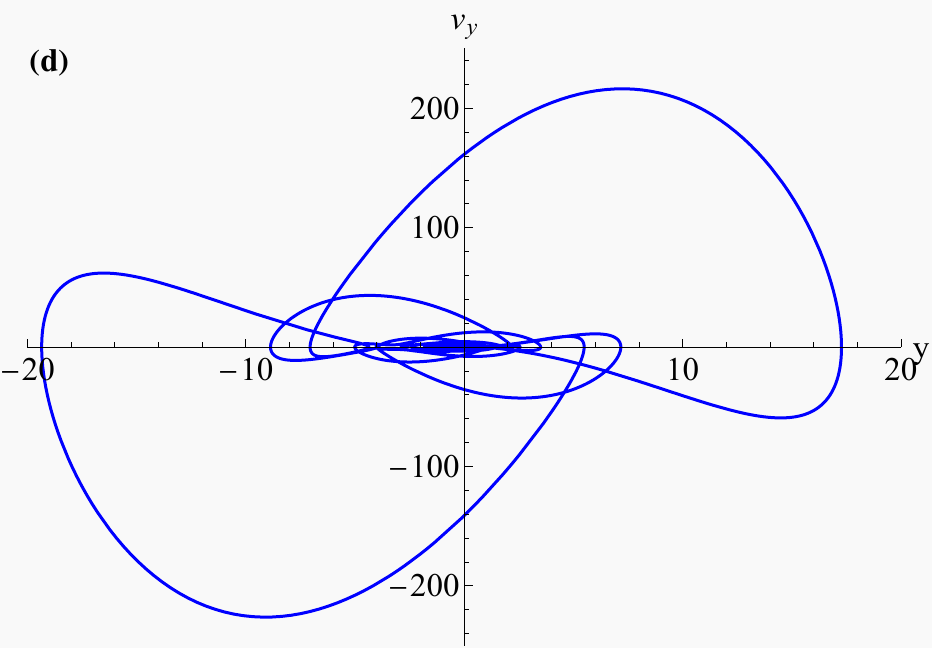}
	\end{minipage}   
	\caption{A nonperiodic trajectory with increasing amplitude for the initial
		conditions
		$x(0)=0.2$, $y(0)=0.4$, $\dot{x}(0)=\dot{y}(0)=0$,
		and parameters
		$\alpha=-\beta=\omega^{2}=1$, $\nu=1$, and $\mu=0.1$,
		over the interval $0\leq t\leq 120$.
		Panel~(a) shows the scalar curl of the force field,	$	\left(\nabla\times\mathbf{F}\right)_{z}	=	\partial_{x}F_{y}-\partial_{y}F_{x},	$
		as a colour map, the force field $\mathbf{F}(x,y)$ as arrows,
		and the resulting trajectory in configuration space in red.
		Panel~(b) shows the corresponding trajectory in velocity space
		$(v_x,v_y)$.
		Panels~(c) and~(d) show the phase-plane projections
		$(x,v_x)$ and $(y,v_y)$, respectively.}
	\label{random}
\end{figure}

We begin by choosing initial conditions for which the motion is
nonperiodic. A representative example is shown in figure \ref{random}. 

In panel~(a), the scalar curl of the force field is
displayed as a colour map, while the arrows represent the force field
$\mathbf{F}(x,y)$ and the resulting trajectory is shown in red. The
initial point is located away from the zero-curl contour. As time
evolves, the trajectory develops successive loops of increasing size,
indicating a growth in the oscillation amplitude.

The arrows should not be interpreted as giving the instantaneous
direction of the trajectory. For a second-order dynamical system, the
tangent to the trajectory is determined by the velocity, whereas the
force determines the acceleration, or more precisely the
position-dependent contribution to it. A force component parallel to
the velocity changes the speed, while a transverse component changes
the direction of motion and bends the trajectory. Consequently, the
red curve may cross the force-field lines rather than remain tangent to
them. The changing curvature of the orbit and the growth of successive
loops reflect the cumulative action of the position-dependent force
field.

\subsection{Periodic motion on the zero-curl line $y=x$}

\begin{figure}[h]
	\begin{minipage}[b]{\textwidth}      
		\centering
		\includegraphics[width=0.49\textwidth]{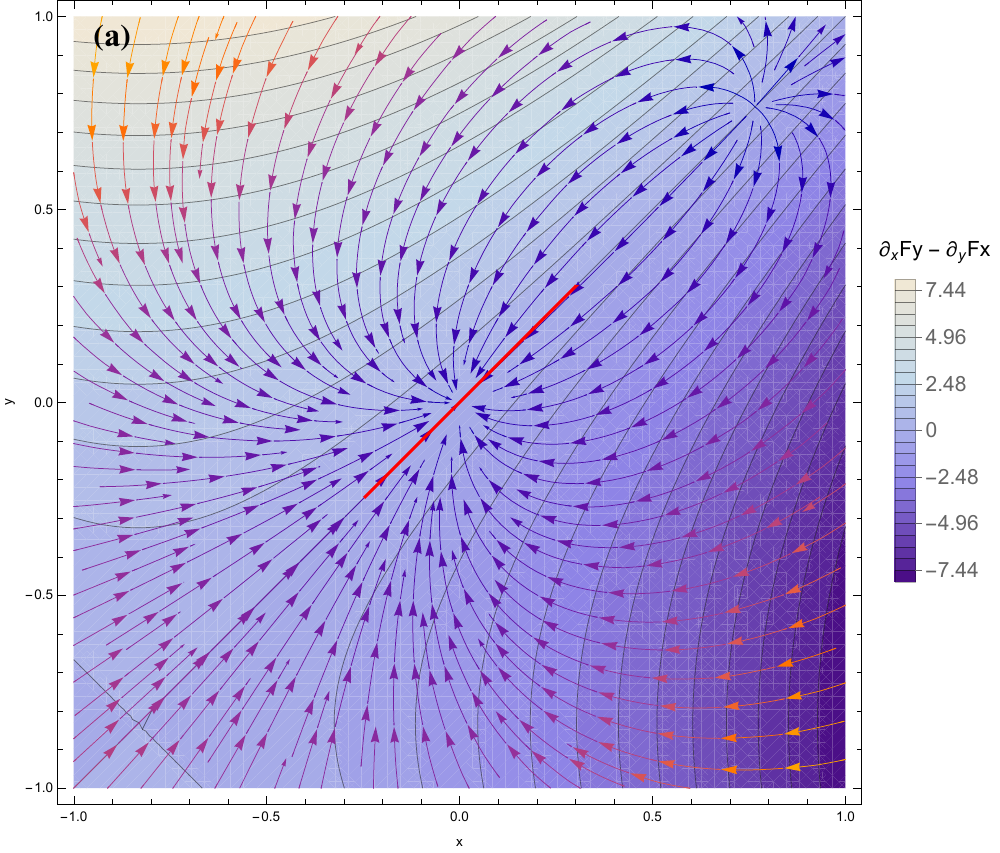}
		\includegraphics[width=0.49\textwidth]{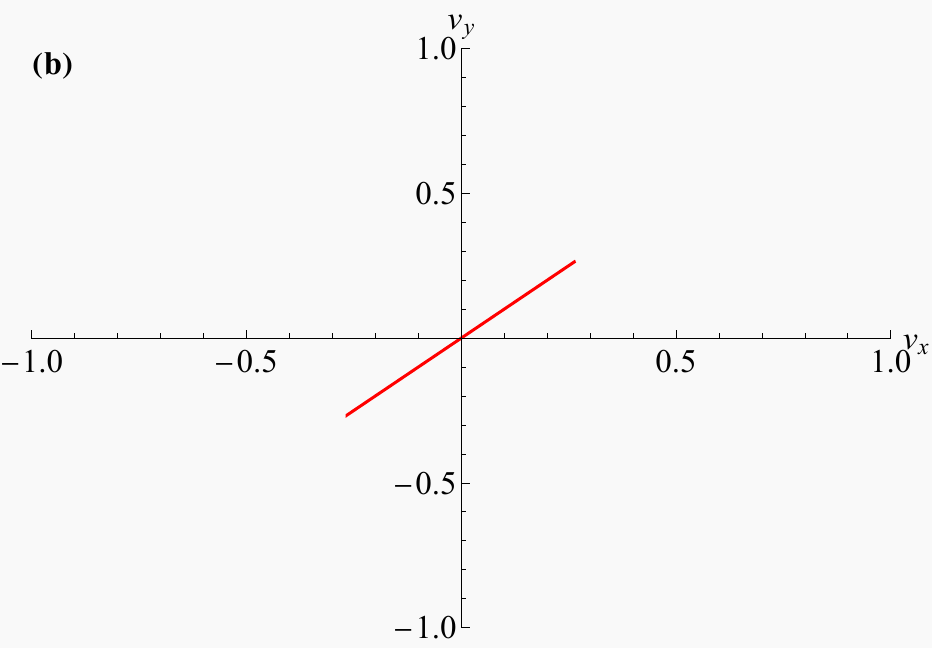}
		\includegraphics[width=0.49\textwidth]{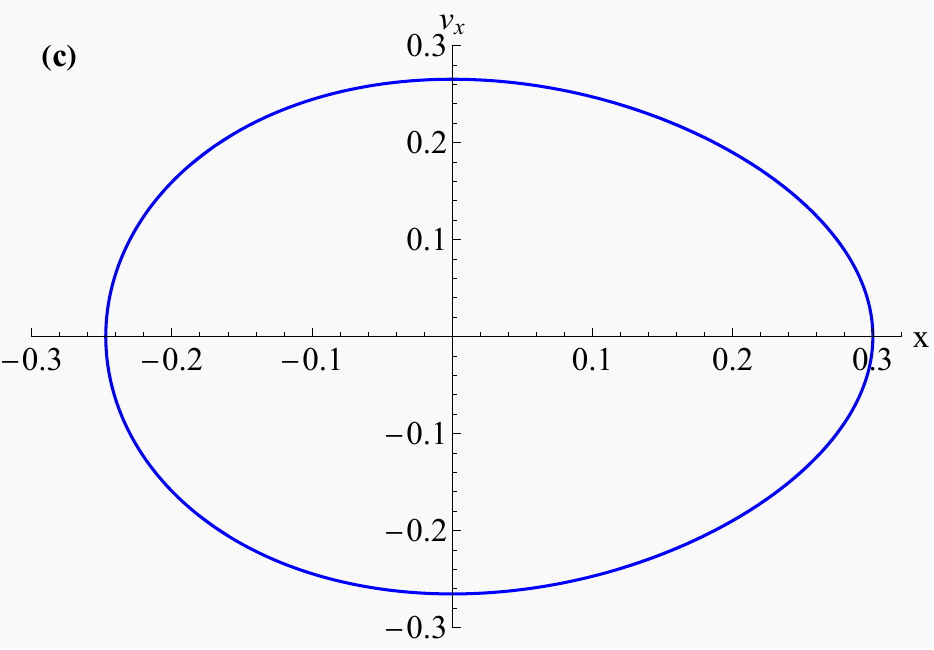}
		\includegraphics[width=0.49\textwidth]{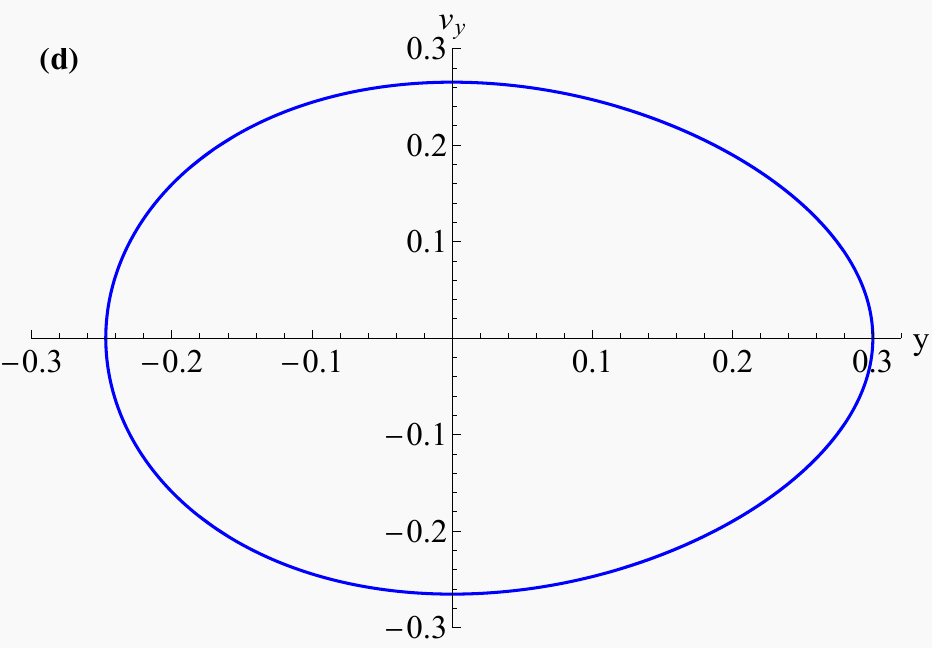}
	\end{minipage}   
	\caption{A periodic trajectory on the zero-curl line $x=y$ with initial conditions
		$x(0)=0.3$, $y(0)=0.3$, $\dot{x}(0)=\dot{y}(0)=0$,
		and parameters
		$\alpha=-\beta=\omega^{2}=1$, $\nu=1/10$, and $\mu=1$,
		over the interval $0\leq t\leq 120$.
		Panel~(a) shows the scalar curl of the force field,	$	\left(\nabla\times\mathbf{F}\right)_{z}	=	\partial_{x}F_{y}-\partial_{y}F_{x},	$
		as a colour map, the force field $\mathbf{F}(x,y)$ as arrows,
		and the resulting trajectory in configuration space in red.
		Panel~(b) shows the corresponding trajectory in velocity space
		$(v_x,v_y)$.
		Panels~(c) and~(d) show the phase-plane projections
		$(x,v_x)$ and $(y,v_y)$, respectively.}
	\label{period1}
\end{figure}

Next, we choose symmetric initial conditions (\ref{zerocurl1}) on the zero-curl line
$y=x$. The resulting dynamics is shown in figure \ref{period1}.

In panel~(a), the trajectory is confined to the diagonal $y=x$ and
oscillates periodically between two turning points. The red segment
therefore represents a path that is repeatedly traversed in opposite
directions, rather than an orbit that closes as a loop in the
$(x,y)$ plane. The force-field arrows along the diagonal are tangent
to this line, so that the acceleration does not generate a component
normal to the invariant manifold. The same symmetry is visible in panel~(b), where the velocity-space
trajectory lies on the line $v_x=v_y$. Panels~(c) and~(d) are
identical closed trajectories, reflecting the relations
$x(t)=y(t)$ and $v_x(t)=v_y(t)$. In contrast to the growing
nonperiodic trajectory discussed above, the amplitudes of both the
position and velocity remain bounded and constant from one cycle to
the next.

Next we determine the exact solution and in particular the period for this motion.  On the invariant manifold associated with the zero-curl line, we set
$x(t)=y(t)=q(t)$. The momentum-velocity relations are determined by the physical $H_1$-flow, $ p_x=\dot{x}=\dot q$, $p_y=-\dot{y}=-\dot q$. The restriction of the first Hamiltonian vanishes identically,
\begin{equation}
H_1\big|_{x=y=q,\;p_x=-p_y}=0.
\end{equation}
The second Hamiltonian, however, remains a nontrivial conserved
quantity. Its restriction to the invariant manifold is
\begin{equation}
H_2
=
-\dot q^{\,2}
-\omega^2q^2
+\frac{2\mu}{3}q^3
+2\nu q^4.
\end{equation}
Setting $H_2=I$, as in (\ref{Hconst}), yields
\begin{equation}
\dot q^{\,2}=P_4(q),
\qquad \text{with} \quad
P_4(q) :=2\nu q^4+\frac{2\mu}{3}q^3-\omega^2q^2-I.
\end{equation}
Differentiating this equation gives
\begin{equation}
\ddot q = \frac12P_4'(q) = \mu q^2+4\nu q^3-\omega^2q,
\end{equation}
which is in agreement with the restriction of the original equations of
motion.

We next derive the exact elliptic-function solution. Assuming $r$ to be a simple root of $P_4$, such that
$ P_4(r)=0$ and $P_4'(r)\neq0$, we introduce the transformation
\begin{equation}
z = \frac{P_4'(r)}{4(q-r)} +\frac{P_4''(r)}{24},
\end{equation}
or equivalently
\begin{equation}
q =r+ \frac{P_4'(r)} {4\left[z-\dfrac{P_4''(r)}{24}\right]},
\end{equation}
so that the first-order equation reduces exactly to the Weierstrass normal form,
\begin{equation}
\dot z^{\,2}=4z^3-g_2z-g_3, \label{Weistrassnf}
\end{equation}
where
\begin{eqnarray}
g_2 &=& \frac{\left(P_4''(r)\right)^2}{48}  -\frac{P_4'(r)P_4'''(r)}{24}=\frac{\omega^4}{12}-2\nu I, \\
g_3 &=& \frac{P_4'(r)P_4''(r)P_4'''(r)}{576}
-\frac{\left(P_4'(r)\right)^2P_4^{(4)}(r)}{384}-\frac{\left(P_4''(r)\right)^3}{1728} =\frac{\omega^6}{216}+\left(\frac{\mu^2}{36}+\frac{\nu\omega^2}{3}\right)I. \quad
\end{eqnarray}
Consequently, a solution of (\ref{Weistrassnf}) is 
\begin{equation}
z(t)=\wp(t-t_0;g_2,g_3),
\end{equation}
where $\wp$ denotes the Weierstrass elliptic functions. The exact position and velocity are therefore
\begin{eqnarray}
q(t) &=&r+ \frac{P_4'(r)} {4\left[	\wp(t-t_0;g_2,g_3)	-\dfrac{P_4''(r)}{24}	\right]},\\
\dot q(t)&=&-\frac{P_4'(r)\wp'(t-t_0;g_2,g_3)}{4\left[	\wp(t-t_0;g_2,g_3)	-\dfrac{P_4''(r)}{24}	\right]^2}.
\end{eqnarray}
The real period of this solution is
\begin{equation}
T=\frac{2K(m)}{\sqrt{e_1-e_3}}, \qquad  \text{with}  \qquad m=\frac{e_2-e_3}{e_1-e_3},    \label{expe}
\end{equation}
where $e_1>e_2>e_3$ are the zeros of the polynomial in the Weierstrass normal form on the right hand side of (\ref{Weistrassnf}), $K(m)$ is the complete elliptic integral of the first kind and $m$ is the square of the elliptic modulus. Although $\wp(t-t_0)$ has a pole at $t=t_0$, the expression
for $q(t)$ has the finite limiting value
$q(t_0)=r$ and $\dot q(t_0)=0$.

In particular, for the initial conditions and parameters used in figure \ref{period1}, $q(0)=0.3$, $\dot q(0)=0$,
$\mu=1$,
$\nu=1/10$,
$\omega^2=1$
we find
\begin{equation}
I=-\frac{3519}{50000},
\qquad
P_4(q) =\frac{q^4}{5}+\frac{2 q^3}{3}-q^2+\frac{3519}{50000}.
\end{equation}
The zeros of $P_4(q)$ are $r_1 \approx -4.45235$, $r_2 \approx -0.247115$, $r_3 = 3/10$ and $r_4 \approx 1.06613$.  The bounded motion therefore occurs between the turning points
$r_2$ and $r_3$. Choosing the initial turning point $r=3/10$ with $t_0=0$, we obtain
\begin{equation}
P_4'(r)=-\frac{249}{625}, \quad     \frac{P_4''(r)}{24}= - \frac{73}{3000}, \quad
g_2=\frac{73057}{750000},
\qquad
g_3=\frac{8873}{27000000}.
\end{equation}
The zeros are found to be $e_1= 0.15771260091909792$, $e_2= -0.003375276573850718$, $e_3= -0.1543373243452472$, so that $m = 0.48377530500452$ and the period according to (\ref{expe}) then becomes 
\begin{equation}
T = 6.5897621034. 
\end{equation}
This value agrees with the period extracted from the numerical solution in figure \ref{period1}.

\subsection{Periodic motion on the zero-curl line $ \mu+6\nu(x+y)=0$}

\begin{figure}[h]
	\begin{minipage}[b]{\textwidth}      
		\centering
		\includegraphics[width=0.49\textwidth]{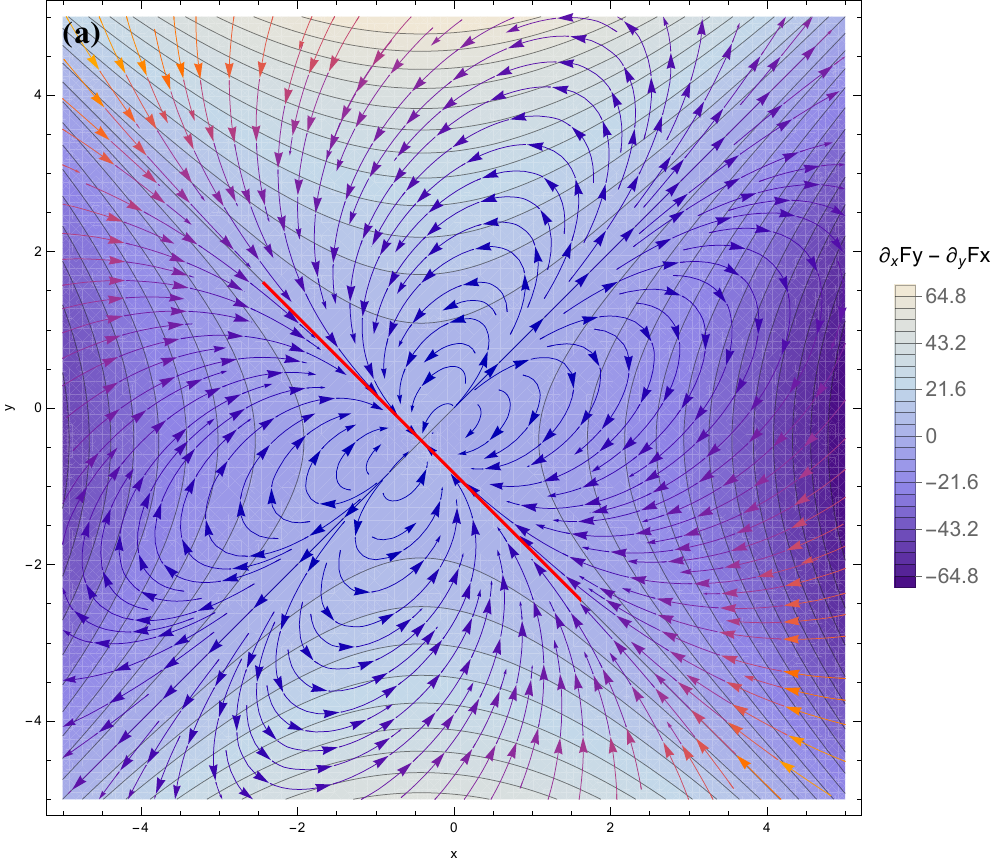}
		\includegraphics[width=0.49\textwidth]{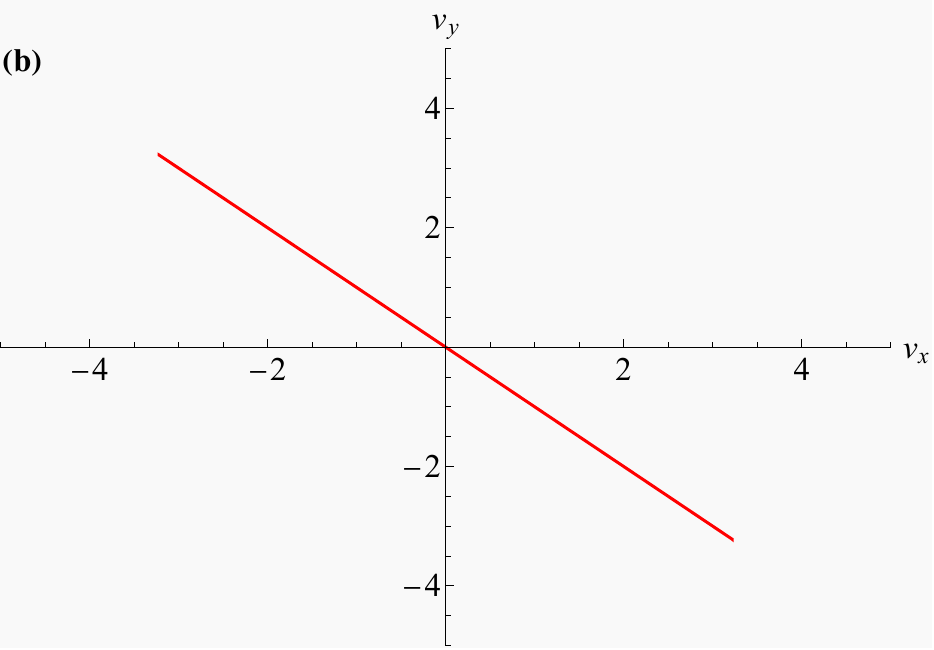}
		\includegraphics[width=0.49\textwidth]{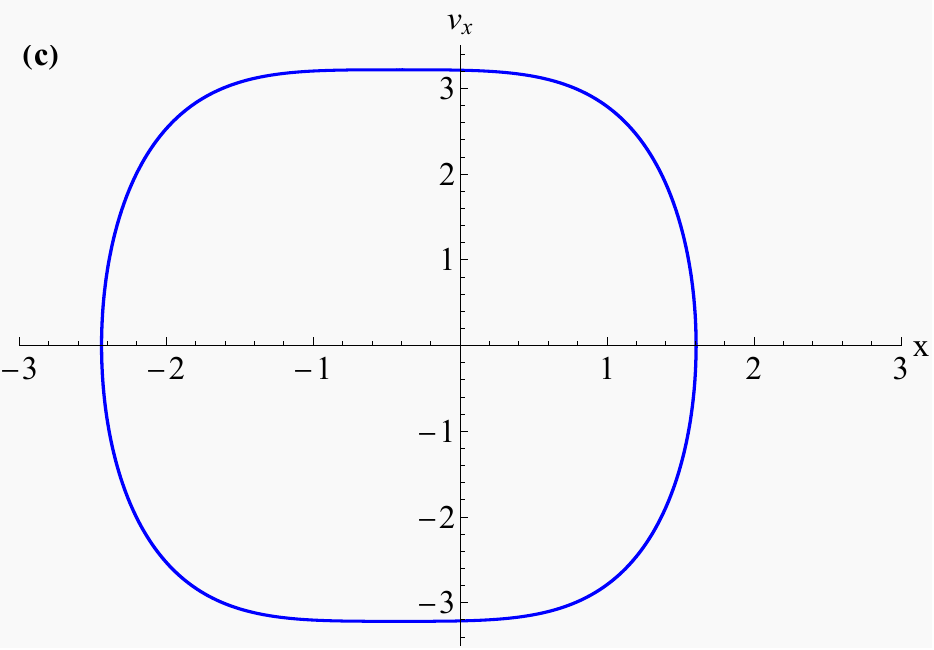}
		\includegraphics[width=0.49\textwidth]{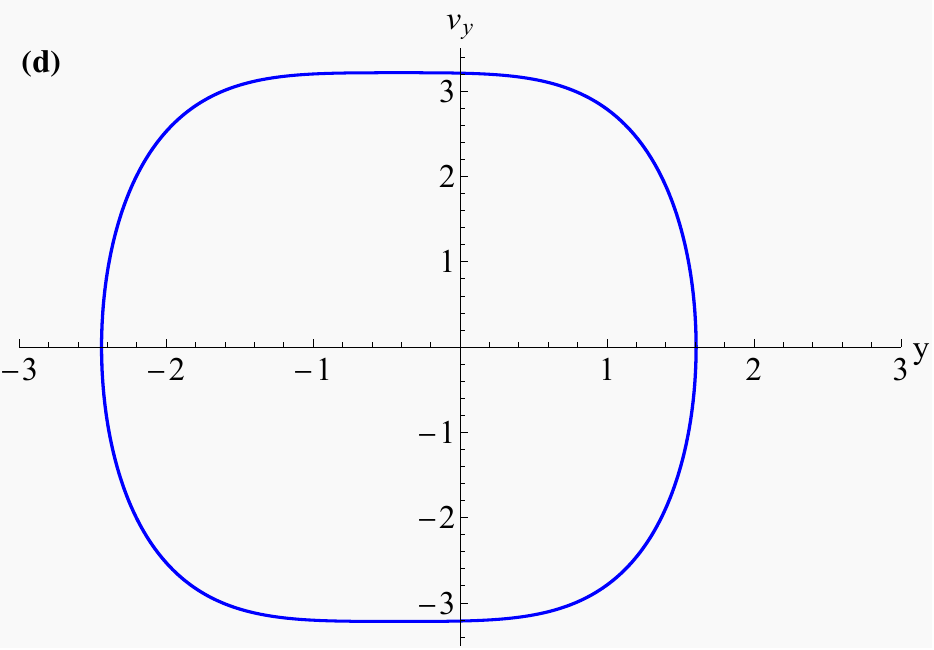}
	\end{minipage}   
	\caption{A periodic trajectory on the zero-curl line $\mu + 6 \nu (x+y)=0$ with initial conditions
		$x(0)=0.2$, $y(0)=- x(0) - \mu/6 \nu $, $\dot{x}(0)=-\dot{y}(0)=3.2$,
		and parameters
		$\alpha=-\beta=\omega^{2}= - \mu^2 / 18 \nu$,  $\nu=1/10$, and $\mu=1/2$,
		over the interval $0\leq t\leq 120$.
		Panel~(a) shows the scalar curl of the force field,	$	\left(\nabla\times\mathbf{F}\right)_{z}	=	\partial_{x}F_{y}-\partial_{y}F_{x},	$
		as a colour map, the force field $\mathbf{F}(x,y)$ as arrows,
		and the resulting trajectory in configuration space in red.
		Panel~(b) shows the corresponding trajectory in velocity space
		$(v_x,v_y)$.
		Panels~(c) and~(d) show the phase-plane projections
		$(x,v_x)$ and $(y,v_y)$, respectively.}
	\label{period2}
\end{figure}

Next, we consider the second zero-curl line $\mu+6\nu(x+y)=0$ with invariant-manifold conditions (\ref{zerocurl2}). The resulting dynamics is shown in figure \ref{period2}.

It is convenient to introduce the coordinate
\begin{equation}
x(t)=-\frac{\mu}{12\nu}+q(t), \qquad   y(t)=-\frac{\mu}{12\nu}-q(t),
\end{equation}
so that
 \begin{equation}  
q(t)=\frac{x(t)-y(t)}{2}, \qquad \dot{x}(t)=\dot q(t), \qquad  \dot{y}(t)=-\dot q(t).
\end{equation}
The restriction of our two bi-Hamiltonians onto this manifold, say $\mathcal M_2$, gives
 \begin{equation} 
H_1\big|_{\mathcal M_2}=0, \qquad H_2\big|_{\mathcal M_2}
=3\dot q^{\,2}+\frac{\mu^4}{10368\nu^3}+\frac{\mu^2}{12\nu}q^2+18\nu q^4.
 \end{equation} 
Setting $H_2=I$ we obtain
\begin{equation}
	\dot q^{\,2}
	=
	C-  a q^2-6\nu q^4, \qquad \text{with} \,\,\, C:=\frac{I-I_0}{3}, \,\,\, I_0:=\frac{\mu^4}{10368\nu^3}, \,\,\, a:=\frac{\mu^2}{36\nu}.   \label{P4se0}
\end{equation}
Differentiating equation (\ref{P4se0}) gives
\begin{equation}
	\ddot q+ a q+12\nu q^3=0.  \label{P4se2}
\end{equation}
Assuming now $q=A$ to be the positive turning point with 
$\dot q=0$ we get
\begin{equation}
	C=aA^2+6\nu A^4, \qquad \Rightarrow \qquad A^2
	=
	\frac{-a+\sqrt{a^2+24\nu C}}{12\nu}. 
\end{equation}
The exact solution of Eq.~(\ref{P4se2}) is then
\begin{equation}
	q(t)	=	A\operatorname{cn}	\left(	\Omega t+\phi_0\mid m	\right),
	\label{JacobiS}
\end{equation}
where $\operatorname{cn}$ is a Jacobi elliptic function and 
\begin{equation}
\Omega^2=a+12\nu A^2, \qquad m=\frac{6\nu A^2}{a+12\nu A^2}.
\end{equation}
The constant $\phi_0$ is determined by the initial position and velocity. Differentiating (\ref{JacobiS}) gives 	
\begin{equation}
	\dot q(t)
	=
	-A\Omega
	\operatorname{sn}
	\left(
	\Omega t+\phi_0\mid m
	\right)
	\operatorname{dn}
	\left(
	\Omega t+\phi_0\mid m
	\right).
\end{equation}
From the properties of the elliptic functions it follows that the exact real period is
\begin{equation}
	T=\frac{4K(m)}{\Omega}.
	\label{PeriodS}
\end{equation}

In particular, for the initial conditions and parameters used in figure \ref{period2}, 
$x(0)= 1/5$, $y(0)=-x(0)- \mu/ 6\nu=-31/30$,
$ \dot{x}(0)=-\dot{y}(0)=16/5$,
$ \mu=1/2$,
$\nu=1/10$,
$\omega^2=-5/36$ and reduced initial conditions 
$ q(0)=37/60$, $\dot q(0)=16/5$.
We find $I = 125815489/ 4050000 \simeq31.0655528395$,  $C= 670885733/64800000$
and $A\simeq2.02397668484$, $\Omega\simeq2.23276115816$,
$m\simeq0.493034970256$. The phase required by the initial conditions is
$\phi_0\simeq-1.41976188802$. 
The period is determined from (\ref{PeriodS}) to
\begin{equation} 
T\simeq3.31109013325.
\end{equation} 
This value agrees with the period extracted from the numerical solution in figure \ref{period2}.

\subsection{An isolated periodic trajectory outside the integrable 	parameter regime}

\begin{figure}[h]
	\begin{minipage}[b]{\textwidth}      
		\centering
		\includegraphics[width=0.49\textwidth]{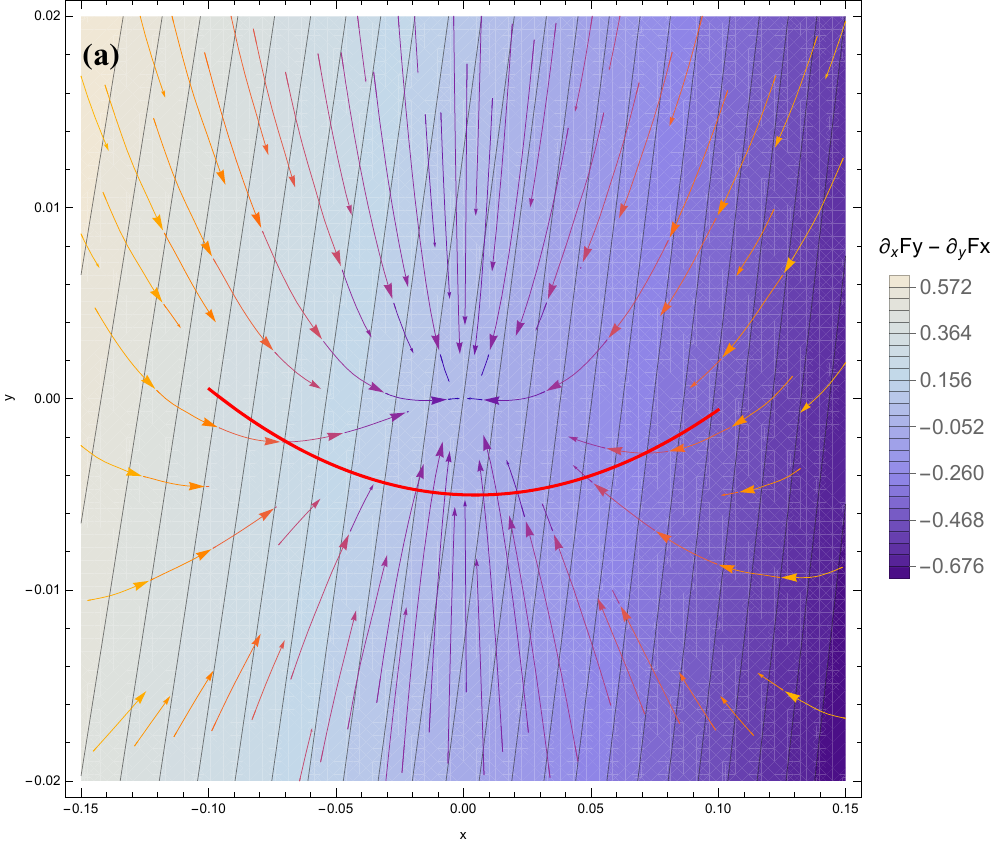}
		\includegraphics[width=0.49\textwidth]{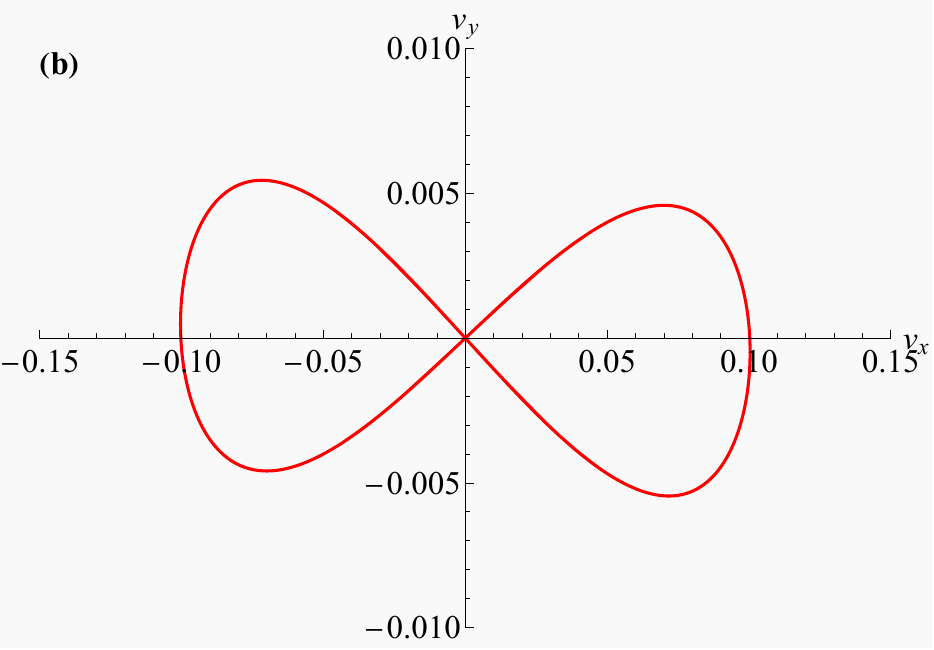}
		\includegraphics[width=0.49\textwidth]{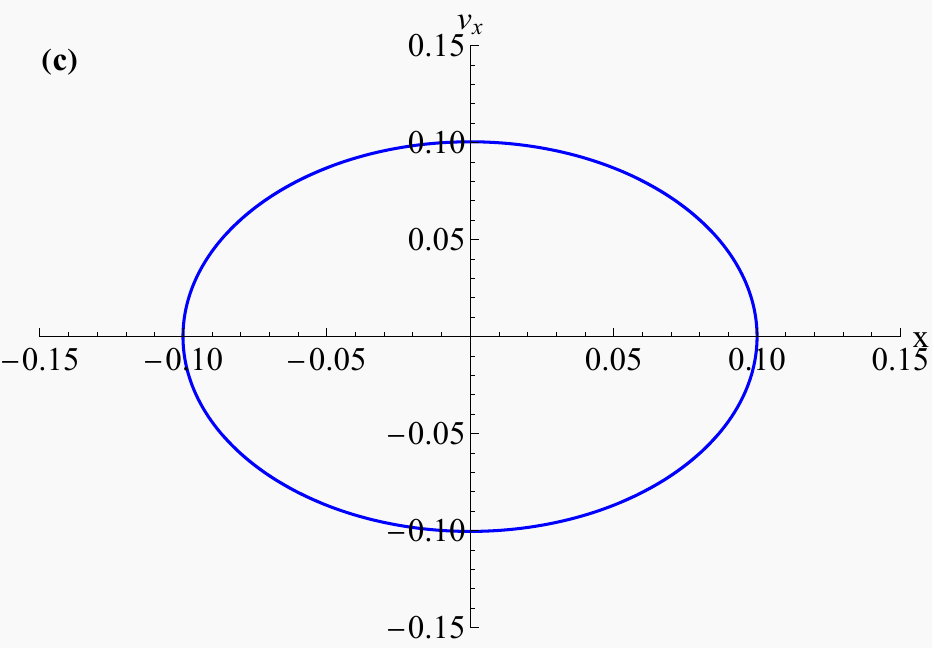}
		\includegraphics[width=0.49\textwidth]{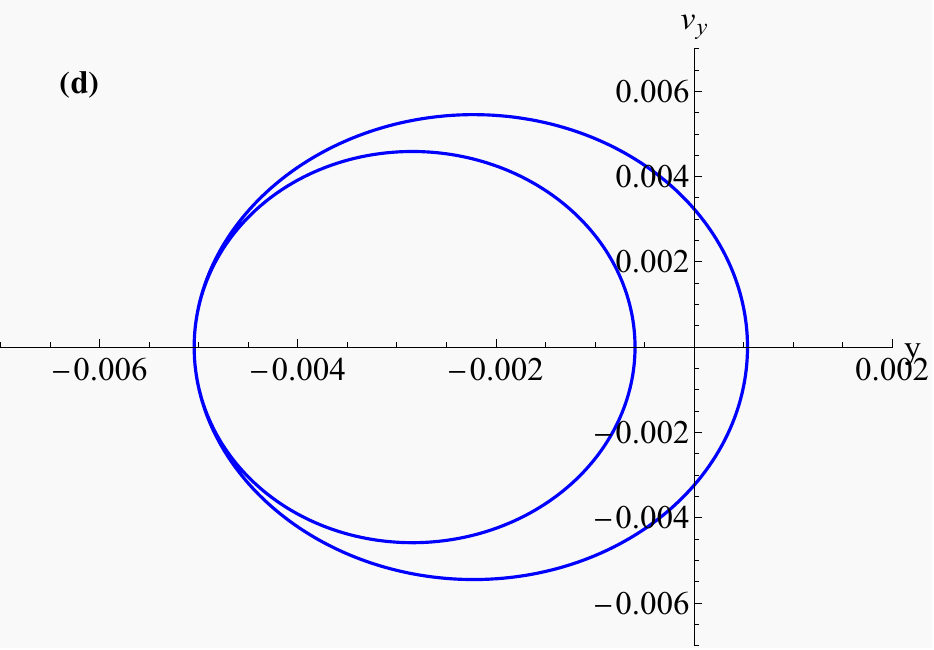}
	\end{minipage}   
	\caption{
		An isolated periodic trajectory, obtained by reversible shooting, in
		the parameter regime $\alpha\neq-\beta$. The initial conditions are
		$x(0)=0.1$, $y(0)=-0.000595082485$, and
		$\dot{x}(0)=\dot{y}(0)=0$, with parameters
		$\alpha=1$, $\beta=-2$, $\mu=1$, and $\nu=1/10$, over the interval
		$0\leq t\leq120$.
		Panel~(a) shows the scalar curl of the force field,
		$(\nabla\times\mathbf{F})_z=\partial_xF_y-\partial_yF_x$, as a
		colour map, the force field $\mathbf{F}(x,y)$ as arrows, and the
		configuration-space trajectory in red. The red curve is traversed
		periodically in both directions.
		Panel~(b) shows the corresponding closed trajectory in velocity space
		$(v_x,v_y)$.
		Panels~(c) and~(d) show the phase-plane projections $(x,v_x)$ and
		$(y,v_y)$, respectively.
	}
	\label{isolatedper}
\end{figure}

We finally demonstrate that the observation of a closed trajectory is
not, by itself, sufficient to infer integrability. For this purpose,
we choose $\alpha=1$, $\beta=-2$, $\mu=1$, $\nu=1/10$. Since $\alpha\neq-\beta$,  the parameter constraint required for the Painlev\'e integrability, the involution of the bi-Hamiltonians and separability structure discussed above is not satisfied. Nevertheless, a periodic trajectory can still be obtained by an appropriate tuning of the initial conditions. The resulting orbit is shown in  figure \ref{isolatedper}.

We exploit now the reversal symmetry (\ref{reversetrans}) to construct periodic orbits. 
We note that if
\begin{equation}
X(t)=(x(t),y(t),v_x(t),v_y(t))
\end{equation}
is a solution, then
\begin{equation}
\mathcal R X(-t),
\qquad
\mathcal R(x,y,v_x,v_y)
=
(x,y,-v_x,-v_y),
\end{equation}
is also a solution. These two solutions need not coincide in general. They do coincide, however, when the initial state lies in the fixed-point set
of $\mathcal R$, namely when $v_x(0)=v_y(0)=0$. In that case, uniqueness implies
$X(t)=\mathcal R X(-t)$, so that the position components are even and the velocity components are odd about $t=0$. This reversibility alone does not guarantee periodicity. However, if the trajectory reaches the fixed-point set again at a time $t_*>0$, so that $v_x(t_*)=v_y(t_*)=0$, then the trajectory is also symmetric about $t=t_*$. It follows that $X(2t_*)=X(0)$, and hence the orbit is periodic with period $
T=2t_*$.

In practice, we fix $x(0)=A$, $\dot{x}(0)=\dot{y}(0)=0$, and regard $y(0)=b$
as a shooting parameter. For each value of $b$, the equations are integrated until the first nontrivial time $t_*(b)>0$ at which $\dot{x}\bigl(t_*(b)\bigr)=0$.
We then define the shooting function $S(b):=\dot{y}\bigl(t_*(b)\bigr)$. A zero of this function $S(b_*)=0$ implies that both velocity components vanish simultaneously at
$t=t_*$. The points at $t=0$ and $t=t_*$ are therefore two turning
points of the motion. By uniqueness of the solution and
time-reversal symmetry,
\begin{equation}
x(t_*+s)=x(t_*-s),
\qquad
y(t_*+s)=y(t_*-s),
\end{equation}
and consequently the full phase-space trajectory returns to its
initial state after the period
\begin{equation}
T=2t_*.
\end{equation}

In our concrete example in figure  \ref{isolatedper} we took $A=0.1$. The shooting procedure gives $b_* =-0.000595082485$. The first simultaneous turning point is reached at $ t_* \simeq3.132794$, where
$x(t_*)\simeq-0.100113348$, $y(t_*)\simeq0.000540019$, $\dot{x}(t_*)=\dot{y}(t_*)=0$. The corresponding numerical period is therefore
$ T_{\mathrm{num}} = 2t_* \simeq6.265588$.

Panel~(a) of figure  \ref{isolatedper} shows that the
configuration-space projection is a curved segment that is traversed
first in one direction and then in the reverse direction. It does not
lie on either of the zero-curl lines. Panel~(b) shows a closed
figure-eight trajectory in velocity space. Panel~(c) shows the nearly elliptical dominant oscillation in the
$(x,v_x)$ plane. Panel~(d) shows a closed two-loop projection in the $(y,v_y)$
plane. The two loops meet tangentially at a common turning point of
the $y$-motion. This self-contact does not violate uniqueness of the
dynamics, since the $(y,v_y)$ plane is only a projection of the full
four-dimensional phase space. In fact, the common projected point is
reached twice during each period with the same values of $x$, $y$,
and $v_y=0$, but with opposite values of $v_x$. The two visits
therefore correspond to distinct full phase-space states. 

For fixed $x(0)=0.1$ and vanishing initial velocities, closure occurs
only after tuning $y(0)$ so that the two velocity components vanish
simultaneously at the second turning point. Nearby values of $y(0)$
give a nonzero shooting function $S(b)$ and do not produce the same
closed trajectory. The orbit is therefore isolated on this
one-dimensional shooting section.

This example illustrates the essential distinction between the
existence of a periodic orbit and Liouville integrability. A periodic
orbit is a one-dimensional invariant set and is a common feature of
both integrable and nonintegrable Hamiltonian systems. By contrast,
Liouville integrability of a two-degree-of-freedom Hamiltonian system
requires a second functionally independent conserved quantity in
involution with the Hamiltonian and, on regular compact levels, a
foliation by invariant tori. The numerical observation of one, or
even several, apparently closed trajectories provides no evidence for
such a global structure. In particular, the isolated periodic orbit
in figure  \ref{isolatedper} exists despite the violation of the
constraint $\alpha=-\beta$ required by the bi-Hamiltonian
construction. It therefore provides a direct warning against
deducing integrability solely from the visual appearance of closed
numerical trajectories.

\section{Higher time-derivative potentialisation and the degenerate PU limit}

The separation of the two-dimensional Hamiltonian system also permits
a direct representation in terms of a higher time-derivative equation.
To construct it, we introduce the differential operator
\begin{equation}
	\mathcal{D}_{\omega}:=
	\frac{d^2}{dt^2}+\omega^2 .
	\label{Domega}
\end{equation}
The potentialisation below is algebraically valid for any real
value of $\omega^2$. When discussing the standard oscillatory PU-limit and the harmonic kernel, however, we restrict to $\omega^2>0$ and take $\omega>0$.
Recalling the separated variables (\ref{uvsep}) the separated equations of motion (\ref{sepeqom}) can be written as
\begin{equation}
	\mathcal{D}_{\omega}u
	+\rho\mu u^2
	-4\bar\rho\nu u^3=0,
	\qquad
	\mathcal{D}_{\omega}v
	+\bar\rho\mu v^2
	-4\rho\nu v^3=0.
	\label{uvHTDTeom}
\end{equation}
For real $x,y,\mu,\nu, \omega^2$, one has $v=\bar u$, so that the two
equations in \eqref{uvHTDTeom} are complex conjugates.

The cube-root phases in these equations can be removed by defining
\begin{equation}
	w:=\rho u=\bar\rho x+\rho y,
	\qquad
	\bar w:=\bar\rho v=\rho x+\bar\rho y.
	\label{wdefinition}
\end{equation}
Using
\begin{equation}
	\rho^3=1,
	\qquad
	\rho+\bar\rho=-1,
	\qquad
	\rho\bar\rho=1,
\end{equation}
together with $u=\bar\rho w$ and $v=\rho\bar w$, the two equations
in \eqref{uvHTDTeom} reduce to
\begin{equation}
	\mathcal{D}_{\omega}w
	+\mu w^2
	-4\nu w^3=0,
	\qquad
	\mathcal{D}_{\omega}\bar w
	+\mu\bar w^2
	-4\nu\bar w^3=0.
	\label{wanharmonic}
\end{equation}
Thus the original real two-dimensional system is equivalently
described by one complex anharmonic oscillator with real
coefficients, together with its complex-conjugate equation.

The transformation \eqref{wdefinition} is invertible. Explicitly we have
\begin{equation}
	x=
	\frac{\bar\rho w-\rho\bar w}{\rho-\bar\rho},
	\qquad
	y=
	\frac{\bar\rho\bar w-\rho w}{\rho-\bar\rho}.
	\label{xyfromw}
\end{equation}
The reality of $x$ and $y$ follows immediately from the condition
that $\bar w$ is the complex conjugate of $w$.

We now introduce a complex higher time-derivative potential $q(t)$
through
\begin{equation}
	w=\mathcal{D}_{\omega}q
	=\ddot q+\omega^2q,
	\qquad
	\bar w=\mathcal{D}_{\omega}\bar q.
	\label{potentialisation}
\end{equation}
Substituting \eqref{potentialisation} into the first equation in
\eqref{wanharmonic} gives the nonlinear fourth-order equation
\begin{equation}
		\mathcal{D}_{\omega}^{\,2}q
		+\mu\left(\mathcal{D}_{\omega}q\right)^2
		-4\nu\left(\mathcal{D}_{\omega}q\right)^3=0 .
	\label{interactingPU}
\end{equation}
Its complex conjugate follows from the second equation in
\eqref{wanharmonic}. In expanded form, equation
\eqref{interactingPU} reads
\begin{equation}
		q^{(4)}
		+2\omega^2\ddot q
		+\omega^4q
		+\mu\left(\ddot q+\omega^2q\right)^2
		-4\nu\left(\ddot q+\omega^2q\right)^3=0 .
	\label{interactingPUexpanded}
\end{equation}
Equation \eqref{interactingPU} therefore provides a natural
interacting higher time-derivative representation of the
integrable curl-force model.

In the absence of the cubic and quartic interactions,
$\mu=\nu=0$, equation \eqref{interactingPU} reduces to
\begin{equation}
	\left(
	\frac{d^2}{dt^2}+\omega^2
	\right)^2q=0,
	\label{degeneratePUlimit}
\end{equation}
which is the equal-frequency, or degenerate, fourth-order
Pais-Uhlenbeck equation \cite{pais1950field}. Writing $q=q_{\rm R}+iq_{\rm I}$, both
real components satisfy
\begin{equation}
	\mathcal{D}_{\omega}^{\,2}q_{\rm R}=0,
	\qquad
	\mathcal{D}_{\omega}^{\,2}q_{\rm I}=0.
\end{equation}
The nonlinear terms in \eqref{interactingPU} may therefore be
viewed as an interaction of the degenerate PU
equation expressed entirely through the combination
$\mathcal{D}_{\omega}q$.

The map from a solution of \eqref{interactingPU} to the original
configuration-space variables is obtained by combining
\eqref{xyfromw} and \eqref{potentialisation},
\begin{equation}
	x=
	\frac{
		\bar\rho\,\mathcal{D}_{\omega}q
		-\rho\,\mathcal{D}_{\omega}\bar q
	}{
		\rho-\bar\rho
	},
	\qquad
	y=
	\frac{
		\bar\rho\,\mathcal{D}_{\omega}\bar q
		-\rho\,\mathcal{D}_{\omega}q
	}{
		\rho-\bar\rho
	}.
	\label{reconstructionHTDT}
\end{equation}
Conversely, given a solution $(x,y)$ of the two-dimensional
Hamiltonian system, one first forms
\begin{equation}
	w=\bar\rho x+\rho y
\end{equation}
and then solves the inhomogeneous oscillator equation
\begin{equation}
	\mathcal{D}_{\omega}q=w.
\end{equation}

An important qualification concerns the kernel of the
potentialisation. The function $q$ is determined by $w$ only up to
the transformation
\begin{equation}
	q\longmapsto q+h,
	\qquad
	\mathcal{D}_{\omega}h=0,
	\label{harmonicredundancy}
\end{equation}
that is
\begin{equation}
	h(t)=A\cos(\omega t)+B\sin(\omega t),
	\qquad
	A,B\in\mathbb{C}.
\end{equation}
Equation \eqref{interactingPU} and the reconstruction formula
\eqref{reconstructionHTDT} are invariant under
\eqref{harmonicredundancy}. The higher time-derivative description
is therefore naturally defined on the equivalence classes
\begin{equation}
	q\sim q+h,
	\qquad
	h\in\ker\mathcal{D}_{\omega}.
	\label{quotientspace}
\end{equation}
A unique representative may, for instance, be selected by imposing
$q(t_0)=\dot q(t_0)=0$ at a fixed reference time $t_0$.

This harmonic redundancy is particularly relevant in the free
limit. Although \eqref{degeneratePUlimit} is precisely the
degenerate PU equation, the original two-dimensional
oscillator is represented by its solution space modulo the
ordinary harmonic modes in $\ker\mathcal{D}_{\omega}$. Consequently,
the construction should be understood as a higher time-derivative
potentialisation rather than as an invertible canonical
transformation between the two phase spaces.

Finally, the complex anharmonic equation in
\eqref{wanharmonic} possesses the conserved quantity
\begin{equation}
	\mathcal{E}_{w}
	=
	\frac{1}{2}\dot w^{\,2}
	+\frac{\omega^2}{2}w^2
	+\frac{\mu}{3}w^3
	-\nu w^4,
	\qquad
	\frac{d\mathcal{E}_{w}}{dt}=0.
	\label{wenergy}
\end{equation}
In terms of the higher time-derivative potential this becomes
\begin{equation}
	\mathcal{E}_{q}
	=
	\frac{1}{2}
	\left[
	\frac{d}{dt}
	\left(\mathcal{D}_{\omega}q\right)
	\right]^2
	+\frac{\omega^2}{2}
	\left(\mathcal{D}_{\omega}q\right)^2
	+\frac{\mu}{3}
	\left(\mathcal{D}_{\omega}q\right)^3
	-\nu
	\left(\mathcal{D}_{\omega}q\right)^4.
	\label{qenergy}
\end{equation}
Its complex conjugate supplies the second separated conserved
quantity. Hence the integrability inherited from the separated
Hamiltonian formulation remains manifest in the higher
time-derivative representation.

\section{Conclusions}

We have investigated Hamiltonian curl-force systems with indefinite kinetic
energy, motivated both by Berry and Shukla's programme on curl forces and
by their close relation to ghostly Hamiltonians and higher time-derivative
dynamics. We first revisited Berry's polynomial Hamiltonian curl-force
example. Although this system exhibits numerically closed trajectories, its
Painlev\'e analysis yields only non-principal dominant balances with
resonance spectrum $r=-1,-1,4,4$. The corresponding Laurent
series cannot accommodate the required four arbitrary constants of the
general solution, and the model therefore fails the standard
Painlev\'e test. We then introduced a four-parameter an integrable polynomial modification which remains a genuine
curl-force Hamiltonian. For this family, the condition
\begin{equation}
\alpha+\beta=0   \label{alpbet}
\end{equation}
emerges from several independent viewpoints. It is the compatibility
condition in the Painlev\'e analysis, the condition for the existence of a
second Hamiltonian, the condition under which the two Hamiltonians form a
bi-Hamiltonian pair, and the condition that allows separation in complex
characteristic variables. On this integrable locus we explicitly constructed
the second Hamiltonian, the compatible Poisson tensors, the separated
one-dimensional systems, and a Lax representation whose spectral invariants
encode the two Hamiltonians.

We also showed that the construction is not restricted to the quartic
example. The separated formulation provides a systematic way of generating
integrable polynomial curl-force Hamiltonians of arbitrary degree. In
addition, the same variables lead naturally to a higher time-derivative
potentialisation, whose free limit is the degenerate Pais-Uhlenbeck
oscillator, which is also reflected in property (\ref{alpbet}). This reinforces the connection between Hamiltonian curl forces,
ghostly systems and higher time-derivative theories.

Finally, we analysed the geometry of the curl field and its implications
for periodic trajectories. The zero-curl lines provide invariant reductions
only under additional tangency and parameter constraints, and on these
invariant lines the motion can be solved explicitly in terms of elliptic functions. We
also constructed an isolated periodic orbit outside the integrable parameter
regime. This demonstrates that the existence of closed trajectories is not, by itself, evidence for Liouville or Painlev\'e integrability. The relevant distinction is between isolated periodic motion
and the global integrable structures exhibited by the deformed model.

The zero-curl analysis also provides a practical diagnostic for ghostly
dynamics. Since classical ghostly systems often contain both periodic and
runaway motions, invariant zero-curl lines help to identify bounded
one-dimensional reductions on which the dynamics can be solved explicitly.
In the degenerate higher time-derivative limit these periodic ghostly
motions may nevertheless be mapped to runaway solutions because the explicit
map from one system to the other, such as a Jordan map, involves the time explicitly. This makes the curl-force perspective
especially useful, as it gives a geometric handle on bounded sectors of the
ghostly Hamiltonian system and clarifies how they are represented in the
associated higher time-derivative theory.

Several natural directions remain open. The separated anharmonic
oscillators provide an immediate starting point for a semiclassical
quantisation by WKB methods. Such an analysis would, however, require a
separate discussion of complex contours, Stokes sectors, boundary
conditions and normalisability, and is therefore beyond the scope of the
present classical study. A related problem is the full quantum theory of
the curl-force Hamiltonians constructed here, where one expects the same
domain, spectral and normalisability issues that occur in ghostly
Hamiltonians and higher time-derivative systems. 

Another natural direction is to study perturbations away from the
integrable locus $\alpha+\beta=0$. Since isolated periodic orbits persist
outside this locus, it would be interesting to analyse their stability,
their bifurcations and the transition from regular to chaotic
nondissipative dynamics. Finally, the separated polynomial construction in
Appendix~A suggests a systematic route to integrable curl-force
Hamiltonians of arbitrary degree. Exploring these higher-degree models,
and their corresponding higher time-derivative potentialisations, may
provide a broader class of examples in which curl-force geometry, ghostly
Hamiltonian dynamics and higher time-derivative theory can be studied in
parallel.

\smallskip

\noindent {\bf Acknowledgements}: AlF is supported by JSPS Postdoctoral Fellowships for Research in Japan facilitated through the Alexander von Humboldt foundation, and by JSPS KAKENHI Grant Number JP26KF0009.

\section*{Appendix}
\appendix

\section{Polynomial solutions of the bi-Hamiltonian compatibility equations}
\label{polynomialpot}

In this appendix, we determine the general polynomial potentials
$V(x,y)$ and $W(x,y)$ satisfying the general compatibility conditions (\ref{VWequn}) and  (\ref{Vequn}) 
\begin{equation}
	V_x+2V_y-W_x =0, \qquad 
	V_y+2V_x+W_y=0, \qquad 
	V_{xx}+V_{xy}+V_{yy}=0,  \label{WVcompat}
\end{equation}
required for the two Hamiltonians $H_1$ and $H_2$ in (\ref{H1V}) and (\ref{H2W}) to be in involution. Given the factorisation (\ref{factoriseop}) we already derived that the solution is of the general form (\ref{Vxyuv}).
Using this, we find
\begin{align}
	V_x+2V_y
	&=
	(\rho+2)F'(u)+(\bar\rho+2)G'(v),\\
	-V_y-2V_x
	&=
	-(1+2\rho)F'(u)
	-(1+2\bar\rho)G'(v),
\end{align}
which by (\ref{WVcompat}) integrates to
\begin{equation}
	W(x,y)
	=
	i\sqrt{3}\left[G(v)-F(u)\right]+W_0,
	\label{WFG}
\end{equation}
where $W_0$ is an arbitrary constant.

For real $x$ and $y$, one has $v=\bar u$. A general real polynomial
pair of degree at most $N$ that satisfies (\ref{Vxyuv}) and (\ref{WFG}) can therefore be written as
\begin{align}
	V(x,y)
	&=
	V_0+
	2\operatorname{Re}
	\left[
	\sum_{n=1}^{N}c_n
	\left(y+\rho x\right)^n
	\right],
	\label{app:Vcomplex}\\
	W(x,y)
	&=
	W_0+
	2\sqrt{3}\operatorname{Im}
	\left[
	\sum_{n=1}^{N}c_n
	\left(y+\rho x\right)^n
	\right],
	\label{app:Wcomplex}
\end{align}
where $V_0,W_0\in\mathbb{R}$ and $c_n\in\mathbb{C}$ are arbitrary.
Thus, at each homogeneous degree $n\geq1$, there are two independent
real polynomial solutions.

For comparison with the potentials used in the main text, it is
convenient to introduce two real homogeneous polynomials $P_n$ and
$Q_n$ at each degree. We choose
\begin{align}
	P_1(x,y)
	&=
	x,
	&
	Q_1(x,y)
	&=
	x-2y,
	\label{PQ1}\\[1ex]
	P_2(x,y)
	&=
	\frac{1}{2}\left(x^2-y^2\right),
	&
	Q_2(x,y)
	&=
	\frac{1}{2}
	\left(x^2-4xy+y^2\right),
	\label{PQ2}\\[1ex]
	P_3(x,y)
	&=
	xy^2-x^2y,
	&
	Q_3(x,y)
	&=
	-\frac{1}{3}
	\left(
	2x^3-3x^2y-3xy^2+2y^3
	\right),
	\label{PQ3}\\[1ex]
	P_4(x,y)
	&=
	x^4-y^4+4xy^3-4x^3y,
	&
	Q_4(x,y)
	&=
	-\left(
	x^4+4x^3y-12x^2y^2
	+4xy^3+y^4
	\right).
	\label{PQ4}
\end{align}
Each of these polynomials satisfies the last equation in (\ref{WVcompat}) and moreover,
 two members of each pair are related by
\begin{equation}
	Q_{n,x}=	P_{n,x}+2P_{n,y},  \qquad 	Q_{n,y}=	-P_{n,y}-2P_{n,x}. \label{PtoQ2}
\end{equation}

The most general real polynomial solution of the last equation in (\ref{WVcompat}) of total degree at most four can be written as
\begin{equation}
	V(x,y)
	=
	V_0+
	\sum_{n=1}^{4}
	\left[
	a_nP_n(x,y)+b_nQ_n(x,y)
	\right],
	\label{app:VgeneralPQ}
\end{equation}
with  constants $V_0,a_n,b_n\in\mathbb{R}$. The corresponding potential $W$ is
\begin{equation}
	W(x,y)
	=
	W_0+
	\sum_{n=1}^{4}
	\left[
	a_nQ_n(x,y)-3b_nP_n(x,y)
	\right],
	\label{app:WgeneralPQ}
\end{equation}
where $W_0$ is an independent additive constant. Writing these expressions explicitly, we obtain
\begin{align}
	V(x,y)
	=& \,\,
	V_0
	+a_1x+b_1(x-2y)+
	\frac{a_2}{2}(x^2-y^2)
	+
	\frac{b_2}{2}(x^2-4xy+y^2) +
	a_3(xy^2-x^2y) \nonumber
	\\
	& -\frac{b_3}{3}
	\left(
	2x^3-3x^2y-3xy^2+2y^3
	\right)+
	a_4
	\left(
	x^4-y^4+4xy^3-4x^3y
	\right)\\
	&-
	b_4
	\left(
	x^4+4x^3y-12x^2y^2
	+4xy^3+y^4
	\right), \nonumber
\end{align}
and
\begin{align}
	W(x,y)
	={}&
	W_0
	+a_1(x-2y)-3b_1x
	+
	\frac{a_2}{2}
	(x^2-4xy+y^2)
	-\frac{3b_2}{2}(x^2-y^2) \qquad \qquad
	\nonumber\\
	&-
	\frac{a_3}{3}
	\left(
	2x^3-3x^2y-3xy^2+2y^3
	\right)
	-3b_3(xy^2-x^2y)
	\\
	&-a_4
	\left(
	x^4+4x^3y-12x^2y^2
	+4xy^3+y^4
	\right)-
	3b_4
	\left(
	x^4-y^4+4xy^3-4x^3y
	\right). \nonumber
\end{align}
For the choices 
\begin{equation}
	a_1=b_1=b_2=b_3=b_4=0,
	\quad
	a_2=\alpha,
	\quad
	a_3=\mu,
	\quad
	a_4=\nu,
	\quad
	V_0=W_0=0,
\end{equation}
we recover precisely the two bi-Hamiltonians $H_1$ and $H_2$ in (\ref{Hamint1b}) and (\ref{Hamint2}).

\newif\ifabfull\abfulltrue


\end{document}